\documentclass[prd, twocolumn, superscriptaddress,floatfix, nofootinbib, preprintnumbers]{revtex4-2}
\usepackage[utf8]{inputenc}
\usepackage[colorlinks=true,citecolor=blue]{hyperref}
\usepackage{amssymb}
\usepackage{amsmath}
\usepackage{graphicx}
\usepackage{footmisc}
\usepackage{url}
\usepackage{multirow}
\usepackage{makecell}
\usepackage{hhline}
\usepackage{comment}
\usepackage{array}
\usepackage{float}
\usepackage{ragged2e}
\usepackage{threeparttable}
\usepackage{enumitem}

\usepackage{xcolor}

\def\beq{\begin{equation}}
\def\eeq{\end{equation}}
\newcommand{\bea}{\begin{eqnarray}\begin{aligned}}
\newcommand{\eea}{\end{aligned}\end{eqnarray}}

\newcommand{\mycomment}[1]{}

\begin{document}

\title{Residual ANODE}

\author{Ranit Das}
\email{ranit@physics.rutgers.edu}
\affiliation{NHETC, Dept.\ of Physics and Astronomy, Rutgers University, Piscataway, NJ 08854, USA}

\author{Gregor Kasieczka}
\email{gregor.kasieczka@uni-hamburg.de}
\affiliation{Institut f\"{u}r Experimentalphysik, Universit\"{a}t Hamburg, 22761 Hamburg, Germany}

\author{David Shih}
\email{shih@physics.rutgers.edu}
\affiliation{NHETC, Dept.\ of Physics and Astronomy, Rutgers University, Piscataway, NJ 08854, USA}

\begin{abstract}

We present \textsc{R-Anode}, a new method for data-driven, model-agnostic resonant anomaly detection that raises the bar for both performance and interpretability. The key to \textsc{R-Anode} is to enhance the inductive bias of the anomaly detection task by fitting a normalizing flow directly to the small and unknown signal component, while holding fixed a background model (also a normalizing flow) learned from sidebands. In doing so, \textsc{R-Anode} is able to outperform all classifier-based, weakly-supervised approaches, as well as the previous \textsc{Anode} method which fit a density estimator to all of the data in the signal region instead of just the signal. We show that the method works equally well whether the unknown signal fraction is learned or fixed, and is even robust to signal fraction misspecification. Finally, with the learned signal model we can sample and gain qualitative insights into the underlying anomaly, which greatly enhances the interpretability of resonant anomaly detection and offers the possibility of simultaneously discovering and characterizing  the new physics that could be hiding in the data.
 
\end{abstract}

\maketitle

\section{Introduction}

Despite countless searches at the LHC~\cite{atlasexoticstwiki,atlassusytwiki,atlashdbspublictwiki,cmsexoticstwiki,cmssusytwiki,cmsb2gtwiki,lhcbtwiki}, so far none have turned up any definitive evidence for new physics beyond the Standard Model yet. Since the vast majority of these searches have been model-specific,  there has been increasing interest \cite{Kasieczka:2021xcg,Aarrestad:2021oeb,Karagiorgi:2022qnh} in developing new, model-agnostic search strategies powered by modern machine learning in recent years. The hope is that these will unlock vast, untapped discovery potential in the LHC data that has been missed by all the model-specific searches so far.

Among the model-agnostic search strategies proposed so far, methods for {\it resonant anomaly detection} -- or using additional features $x$ to enhance the sensitivity of a bump hunt in a primary resonant variable $m$  -- have received a lot of attention~\cite{Collins:2018epr,Collins:2019jip,anode,Andreassen:2020nkr,Stein:2020rou,Amram:2020ykb,cathode,Collins:2021nxn,1815227,Kasieczka:2021tew,Hallin:2022eoq,Chen:2022suv,Kamenik:2022qxs,Sengupta:2023xqy,Raine:2022hht,Golling:2023yjq, feta,bickendorf2023combining, cathodebdt,anode_bdt,full_AD}. All of these methods have focused on learning good approximations to the Neyman-Pearson optimal anomaly detector:
\begin{equation}
\label{eq:Roptimal}
R_{\rm optimal}(x) = {p_{\rm data}(x)\over p_{\rm bg}(x)}
\end{equation}
Here $x$ are the additional features, and $p_{\rm data}(x)$ and $p_{\rm bg}(x)$ are the data and background probability densities in the signal region (SR), defined as a window in $m$. Cutting on $R_{\rm optimal}(x)$ can potentially enhance the significance of any anomaly in the signal region by a large amount. 

The main strategy for learning $R_{\rm optimal}$ has been based on weak supervision: training a classifier between the data and a sample of events constructed to resemble the background as closely as possible. One exception to this has been \textsc{Anode} \cite{anode}, which is technically an {\it unsupervised} approach. With \textsc{Anode}, one trains conditional density estimators (normalizing flows in practice) on the signal region and sideband events, interpolates the latter into the SR, and constructs $R_{\rm optimal}$ by taking their ratio directly. Since this approach is solely based on unsupervised density estimation and does not involve any classifiers, it is technically an unsupervised approach to resonant anomaly detection. 

It has been recognized \cite{anode,cathode} that since density estimation is much more difficult than classification, \textsc{Anode} suffers in sensitivity relative to classifier-based approaches. However, weak supervision and classifier-based approaches also have their drawbacks. In particular they do not perform well when the number of signal events is too small and they can be confused by noisy or uninformative features. Finally, the classifier by itself is not interpretable in that it does not tell us where the signal is, only where it is more overdense relative to the background.  

In this paper, we remedy the sensitivity deficit of \textsc{Anode} with a new twist on the method, which we will call {\it residual} \textsc{Anode} or \textsc{R-Anode} for short. The idea of \textsc{R-Anode} is to explicitly fit to the signal component in the SR while holding the background fixed. Assuming data to be a mixture of the signal and background distributions
\begin{equation}\label{eq:jointfit}
p_{\rm data}(x,m) = w\, p_{\rm sig}(x,m) + (1-w)\, p_{\rm bg}(x,m),
\end{equation}
with $w$ as the fraction of signal in the data, and  $p_{\rm bg}(x,m)$ as the fixed background template, we attempt to model $p_{\rm sig}(x,m)$.  

Fitting for the signal density directly, instead of the full data density, not only improves on the performance of \textsc{Anode}, but it even surpasses the performance of the Idealized Anomaly Detector (IAD), which sets an upper bound to classifier-based approaches. The IAD was introduced in \cite{cathode} and corresponds to training a classifier on data vs.\ perfectly simulated background (i.e.\ the same background model that the background in the data was drawn from). The IAD is limited by finite training statistics, noisy features and finite model capacity. So it cannot fully approach the optimal anomaly detector or (by extension) the fully supervised classifier. \textsc{R-Anode} is able to surpass the IAD since it assumes more about the signal vs.\ background mixture.

In \textsc{R-Anode}, one has the option to fix the signal fraction $w$ during training or to let it be a learnable parameter along with the signal density. We explore both options in this work, finding that for fixed $w$, the method is quite robust to $w$-misspecification, retaining excellent sensitivity to the signal even when $w$ is  larger or smaller than the true $w$ by nearly an order of magnitude. For learnable $w$, we find that the method is robust and there is only a slight drop in overall signal sensitivity; furthermore, the learned $w$ tracks the true $w$ well down to a lower threshold of $\sim 200$ signal events. Thus \textsc{R-Anode} could potentially be used to place measure or place limits on the signal fraction directly.

Finally, with the learned signal density, one can also draw potentially unbiased signal samples and directly learn about the properties of the new physics model hiding in the data. In this way, \textsc{R-Anode} simultaneously offers significantly improved performance over other methods and also much greater interpretability.

This paper is organized as follows: Section \ref{sec:ranode} describes the \textsc{R-Anode} method; Section \ref{sec:setup} introduces the datasets and model definitions used in this work; Section \ref{sec:application} describes our results,
including training with fixed $w=w_{\rm true}$ (Sec.~\ref{sec:correct_w}), scanning over fixed $w$ (Sec.~\ref{sec:w_scan}), training with learnable $w$ (Sec.~\ref{sec:learnable_w}), and sampling from the learned signal model (Sec.~\ref{sec:samples}). Finally, Section~\ref{sec:conclusions} contains our conclusions and outlook. Appendix~\ref{appendix:implementation} describes further details of the architecture and hyperparameters used in our implementation of \textsc{R-Anode}, and Appendix~\ref{appendix:undetected} contains more details about the learnable $w$ case.

\section{Residual \textsc{Anode} method}\label{sec:ranode}

In \textsc{R-Anode} we model the signal distribution $p_{\rm sig}(x,m)$ for $m\in {\rm SR}$ directly with a normalizing flow and use it to fit to data from equation (\ref{eq:jointfit}). We minimize the negative log likelihood averaged over the SR data:
\begin{equation}\label{eq:loss} 
L = -{\mathbb E}_{x,m\sim {\rm SR\,\, data}}\log p_{\rm data}(x,m)
\end{equation} 
with respect to the parameters of $p_{\rm sig}(x,m)$ while keeping $p_{\rm bg}(x,m)$ fixed during training. 

To obtain the joint background density in the SR from the sidebands, we break it up into two factors:
\begin{equation}
p_{\rm bg}(x,m) = p_{\rm bg}(x|m) p_{\rm bg}(m)
\end{equation}
The first factor, the  conditional density
$p_{\rm bg}(x|m \in {\rm SR})$, is obtained by interpolating from the sidebands similar to \cite{anode,cathode}. The second factor, the background mass distribution $p_{\rm bg}(m \in {\rm SR})$, can be similarly obtained by interpolating from sidebands or (as we do in this work) approximating it with the data mass distribution under the assumption that there is no statistically significant anomaly in the inclusive bump hunt. This allows us to get $p_{\rm bg}(x,m)$ for $m\in{\rm SR}$. 

For the signal fraction $w$, we explore two options: 
\begin{enumerate}
    \item Hold $w$ fixed during training. We then scan over different values of $w$, exploring the effect of fixing $w$ to be larger or smaller than the true $w$. 
    \item Keep $w$ as a learnable parameter during training, with the same optimizer and hyperparameters used for $p_{\rm sig}(x,m)$.\footnote{Using a different optimizer and hyperparameter to learn $w$ might be interesting, but it could also necessitate further hyperparameter tuning, so we save this for future work.}
    
\end{enumerate} 
We note that the true signal fraction $w_{\rm true}$ is related to the number of signal and background events in the SR:
\begin{equation}
w_{\rm true}={N_{\rm sig,\,SR}\over N_{\rm sig,\,SR}+N_{\rm bg,\,SR}}
\end{equation}
For small amounts of signal that we assume throughout this work, this relation is approximately linear, i.e.\ $w_{\rm true}\approx N_{\rm sig,\,SR}/N_{\rm bg,\,SR}$. 

Finally, the anomaly score is constructed as \begin{equation}
R(x,m) = \frac{p_{\rm sig}(x,m)}{p_{\rm bg}(x,m)}.
\end{equation}

\section{Setup}
\label{sec:setup}

\subsection{Dataset}\label{sec:dataset}

We use the LHCO R\&D Dataset \cite{lhco, Kasieczka:2021xcg} for our studies with 
dataset and train-val-test splits similar to \cite{cathode,anode}. In the following, a brief summary of the dataset is given.

QCD dijet events form the Standard Model(SM) background, and $W' \rightarrow X(\rightarrow qq) Y(\rightarrow qq)$ events with $m_{W'}=3.5\,\text{TeV}$, $m_{X}=500\,\text{GeV}$ and $m_{Y}=100\,\text{GeV}$ are used as signal. These are simulated using \texttt{Pythia 8} \cite{pythia_1,pythia_2} and \texttt{Delphes 3.4.1} \cite{delphes_1, delphes_2, delphes_3}. The reconstructed particles are clustered into jets using the anti-$k_T$ algorithm \cite{antikt_1, antikt_2} with $R=1$ using \texttt{Fastjet} \cite{fastjet}. Events are required to satisfy the $p_T > 1.2\,\text{TeV}$ jet trigger. 

The training features used are 
\begin{equation}
 m=m_{JJ},\quad    x =[m_{J_1}, \Delta m_J, {\tau_{21}}^{J_1},{\tau_{21}}^{J_2}],
\end{equation} 
where invariant masses of the subjets satisfy $m_{J_1}<m_{J_2}$, and $\Delta
m_J = m_{J_2}-m_{J_1}$. The n-subjettiness ratios are defined as $\tau_{ij} = \tau_i/\tau_j$ \cite{nsub_1,nsub_2}. \footnote{ These features correspond to the baseline features defined in \cite{cathodebdt}, along with the invariant mass $m_{JJ}$. We leave exploring the effects of noisy features, and extended set of features for a future study.} The resonant variable is chosen as the dijet invariant mass $m_{JJ}$, with the signal region (SR) defined as $m \in [3.3,3.7]$ and its complement $m \notin [3.3,3.7]$ forming the sideband (SB) regions. 

For the primary dataset (meant to represent actual unlabeled data from the experiment), we use all $1$-million SM background events from the original R\&D dataset and inject different amounts of signal (1000 or lower) from the first 70k signal events of the R\&D dataset. The 
amount of signal in the SR is approximately 76\% of the total injected signal, i.e.\ $N_{\rm sig,\,SR}\approx0.76\times N_{\rm sig}$. Meanwhile, the amount of background in the SR is $N_{\rm bg,\,SR}\approx 120,000$. 
The highest signal-injection of $N_{\rm sig}=1000$ corresponds to $\approx 2\sigma$ significance for the inclusive bump hunt in $m_{JJ}$ (so, below discovery threshold).

All methods split this primary dataset into training and validation sets, with 80-20 split for \textsc{R-Anode}, and 50-50 split for IAD and \textsc{Anode}.\footnote{Following previous work, we use the 50-50 split for ANODE and IAD. For \textsc{R-Anode}, we found that the 80-20 split leads to a better performance  as compared to the 50-50 split.} For IAD, following \cite{cathode}, we used an additional 272k QCD dijet events in the SR \cite{extra_qcd} (with the same split as for data) to train the classifier.  
For evaluating SIC curves, we use the remaining 30k signal events from the original R\&D dataset, along with the additional 340k QCD dijet events in the SR from \cite{extra_qcd}. 

For each $N_{\rm sig}$, ten different datasets are produced by injecting different randomly selected signal events from the R\&D dataset. These are used to derive error bars on the performance curves in the figures below.

\subsection{Model architectures and selection}
\label{sec:model}

Here we briefly describe the model architectures and model selection procedures used in this work. More details can be found in Appendix \ref{appendix:implementation}.

 In \cite{anode,cathode}, the conditional normalizing flow models used to fit the data in the SR and SB regions were 
 Masked Autoregressive Flows (MAF) with affine transformations \cite{maf,nice}. Here we use the same model for learning $p_{\rm bg}(x|m)$ in the SB, but for learning $p_{\rm data}(x|m)$ and $p_{\rm sig}(x,m)$ in the SR (for \textsc{Anode} and \textsc{R-Anode}, respectively), we switch to RQS transformations~\cite{rqs_paper} instead. Using the more expressive RQS transformations for the data and signal distributions was found to improve the performance of the methods. Interestingly, using RQS transformations for fitting the background in the SB led to {\it worse} results. Since RQS transformations are more expressive, it possibly leads to overfitting of the simpler background and/or interpolation.

 For a proof of concept, $p_{\rm bg}(m\in{\rm SR})$ was estimated using the mass histograms of each full dataset using \texttt{scipy} $\texttt{rv\_histogram}$. 
 Even though the full data contains signal, this should be an excellent approximation to $p_{\rm bg}(m)$ under our assumption that there is no significant excess in the inclusive bump hunt. In future work it would be interesting to also compare this against a proper interpolation of the mass histogram from the sidebands. 
 
 Finally, for the IAD and fully supervised classifier, we utilize boosted decision trees (specifically a \texttt{HistGradientBoostingClassifier} from \texttt{scikit-learn} \cite{scikit, lightgbm}), which were shown in \cite{cathodebdt} to be robust under uninformative features.  

For \textsc{Anode} and \textsc{R-Anode}, the method is trained on 20 random train/val splits of a given dataset, and for each split the 10 epochs with lowest validation losses are selected. Probabilities from these 200 models (or fewer in the case of learnable $w$, see Sec.~\ref{sec:learnable_w} for details) are ensembled to obtain the final predictions of the methods. For the IAD, the model is retrained on 50 random train/val splits of a given dataset, and the models with lowest validation loss for each retraining are chosen to ensemble, similar to \cite{cathodebdt}.

\subsection{Performance metrics}

The main performance metric we use in this paper is the Significance Improvement Characteristic (SIC). In terms of the signal efficiency ($\epsilon_S$ or TPR) and background efficiency ($\epsilon_B$ or FPR), one has 
\begin{equation}
\text{SIC} = {\epsilon_S\over\sqrt{\epsilon_B}}
\end{equation}
The SIC characterizes the improvement to the nominal significance, using a cut on the anomaly score from a given method. We will also quote this nominal significance below, which is given by
\begin{equation}
\text{significance} = \text{SIC} \times {N_{\text{sig,\,SR}}\over\sqrt{N_{\text{bg,\,SR}}}}
\end{equation}
The latter factor is the nominal significance from the inclusive bump hunt (i.e.\ prior to any enhancement from a resonant anomaly detection method).

In general, the SIC depends on the working point, and we will exhibit this dependence by plotting SIC vs.\ the signal efficiency.
At lower signal efficiencies, there are large statistical uncertainties in the SIC values. Hence we introduce a cut-off in the SIC curves when the relative statistical error on the background efficiency exceeds 20\%. 

It is also useful to quantify the performance at particular working points. Max SIC is defined as the maximum of all the SIC values above the cut-off for statistical stability. We also use SIC at  $\text{FPR}=0.001$ to represent a typical, fixed working point that an experiment might choose in practice.

\begin{figure*}
    \centering
    \includegraphics[width=0.48\textwidth]{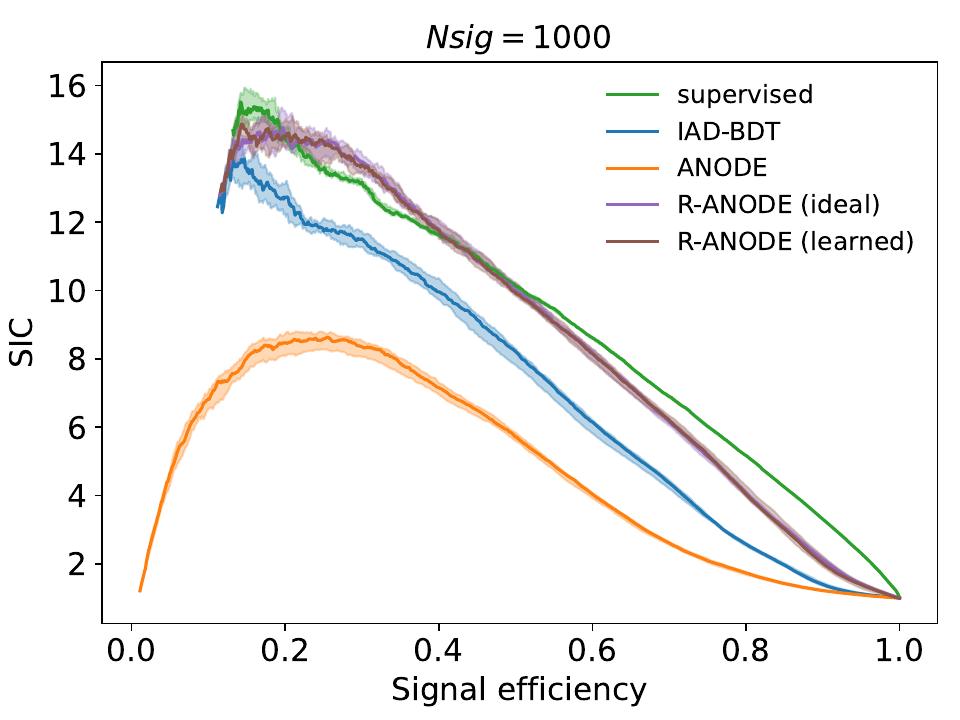}
    \includegraphics[width=0.48\textwidth]{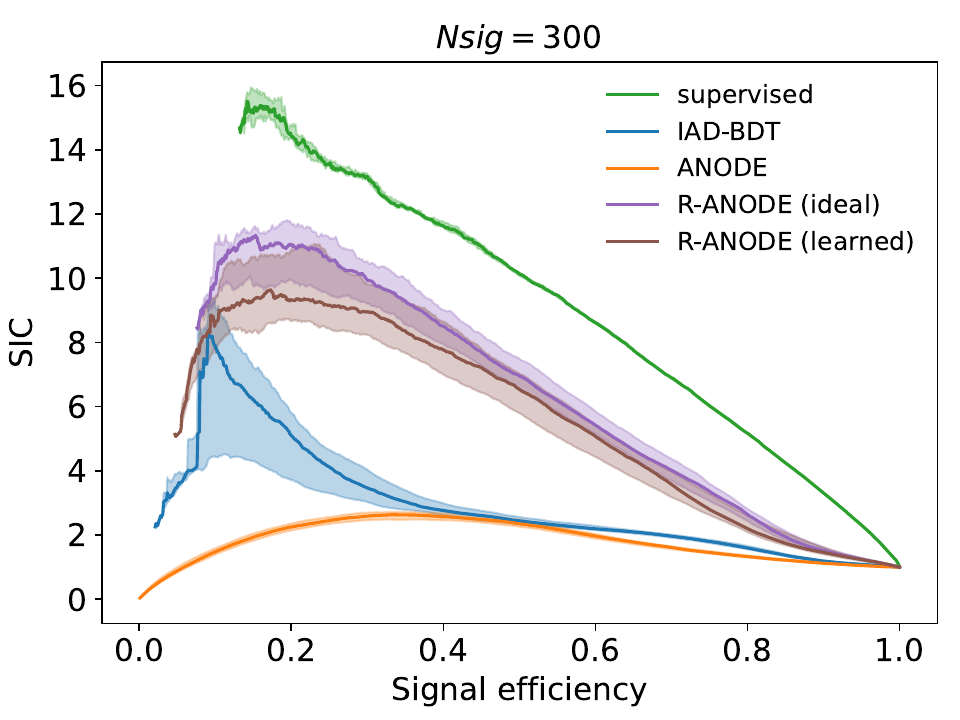}
    \caption{SIC curves for $N_{\rm sig}=1000$ (left) and $N_{\rm sig}=300$ (right). \textsc{R-Anode} with a learnable $w$ almost saturates the performance of an idealized \textsc{R-Anode} where $w$ is fixed to its true value during training. Both these methods outperform the IAD and \textsc{Anode}. The supervised classifier sets the upper limit for performance of all these methods, and at larger signal strengths, \textsc{R-Anode} saturates this upper limit before the IAD does.}
    \label{fig:performance_curves}
\end{figure*}

\begin{figure*}
    \centering
    \includegraphics[width=0.48\textwidth]{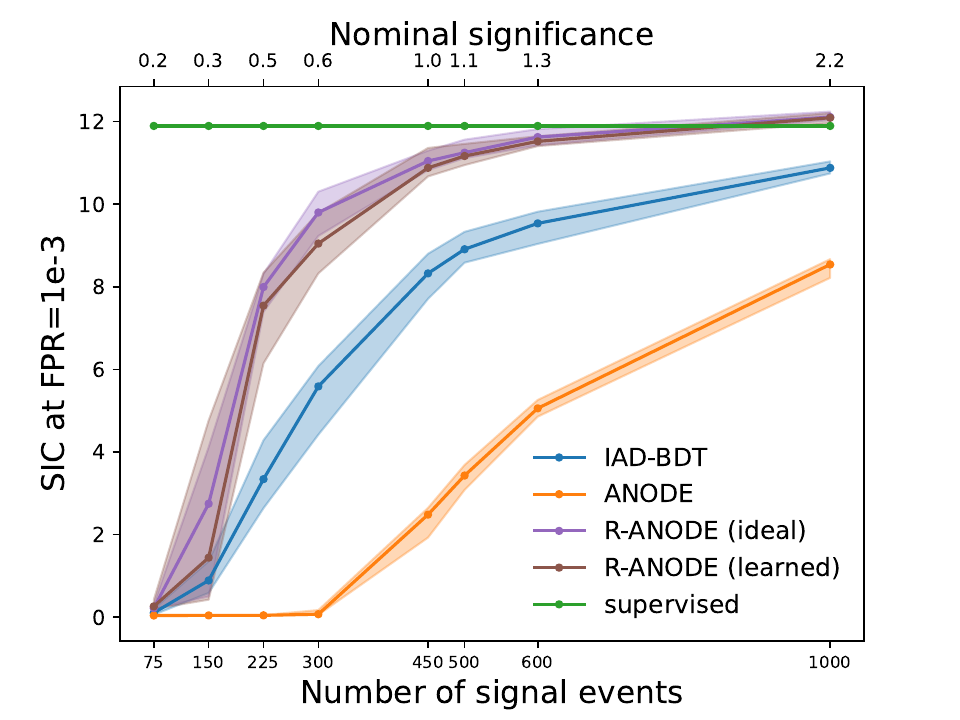}
    \includegraphics[width=0.48\textwidth]{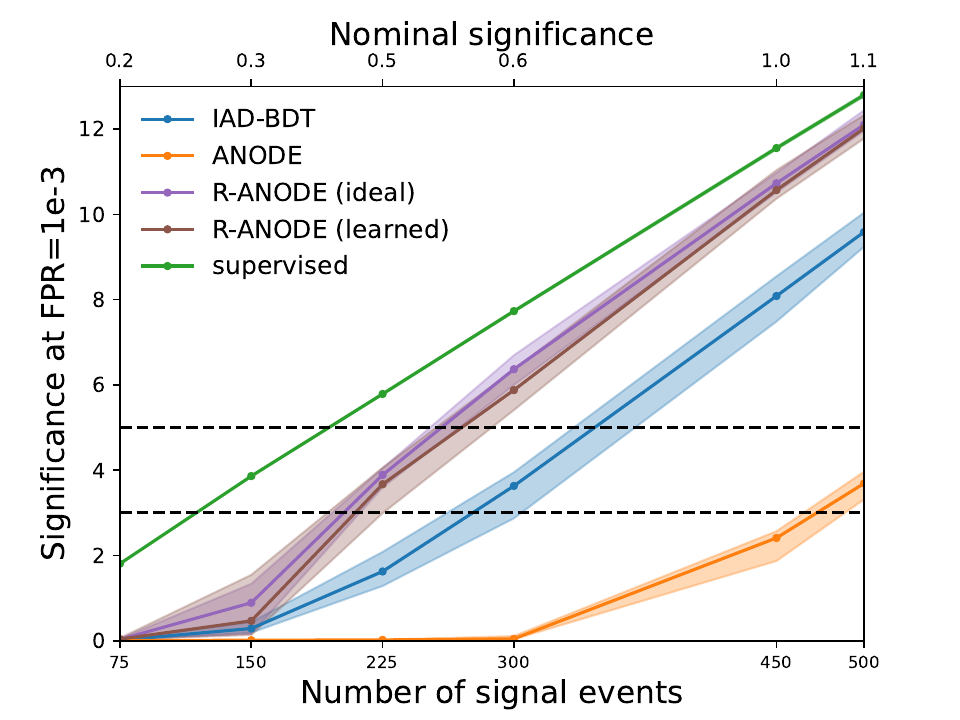}  
    \caption{Left: SIC (at $ \rm FPR = 0.001$) vs $N_{\rm sig}$ (amount of signal injected to data). Right: Total significance achieved (at FPR=0.001) vs $N_{\rm sig}$.  Again we see that \textsc{R-Anode} with learnable $w$ matches the idealized \textsc{R-Anode} with $w=w_{\rm true}$, and both outperform IAD and \textsc{Anode}, across a wide range of signal levels.}
    \label{fig:sic_and_significance}
\end{figure*}

\section{Results}\label{sec:application}

We first show the performance of an ideal version of \textsc{R-Anode} with the correct value of $w=w_{\rm true}$ fixed during training, for different amounts of injected signal. 
Then we present the results for different fixed values of $w$ for $N_{\rm sig}=1000$. And finally, we show the results for the case where we attempt to learn $w$.

We compare \textsc{R-Anode} to \textsc{Anode}, the IAD, and a fully supervised classifier. For all figures, unless otherwise mentioned, the curves show the median value and $68\%$ confidence bands for the results obtained by retraining the methods on 10 different datasets described in Sec.~\ref{sec:dataset} corresponding to different randomly-selected signal injections. Since the upper limit for performance of the classifier-based data-driven approaches like \textsc{CWoLA} \cite{Collins:2018epr,Collins:2019jip}, \textsc{CATHODE} \cite{cathode}, \textsc{CURTAINS} \cite{Raine:2022hht},  etc.\ is the IAD, we omit the explicit comparison to these methods. The supervised classifier sets the upper bound to performance for all methods on this signal hypothesis. Interestingly, as observed in \cite{cathode, cathodebdt, Raine:2022hht}, for these signal injections there is a difference in performance between the IAD and supervised classifier. This is because the IAD is not actually fully optimal -- it is limited by finite training statistics and finite model capacity. The truly optimal AD given by Eq.~(\ref{eq:Roptimal}) would be completely equivalent to the fully supervised classifier, since it would be monotonic with it \cite{cathode}.

\subsection{Idealized version: fixing $w=w_{\rm true}$ }\label{sec:correct_w}

We study the performance of \textsc{R-Anode} for different amounts of signal-injections, in the idealized scenario where $w=w_{\rm true}$ is held fixed during training.  In Fig.~\ref{fig:performance_curves}, we show the SIC curves for $N_{\rm sig}=1000$ and $N_{\rm sig}=300$ (which correspond to 2.2$\sigma$ and 0.7$\sigma$ nominal inclusive significance in the SR, respectively). In both cases, the \textsc{R-Anode} (ideal) method outperforms \textsc{Anode} and the IAD across a wide range of signal efficiencies. 
For $N_{\rm sig}=1000$, \textsc{R-Anode} (ideal) nearly closes the gap between the IAD and the fully supervised classifier\footnote{Interestingly, \textsc{R-Anode} even seems to outperform the fully supervised classifier for a small range of  signal efficiencies. This shouldn't be possible and it may not be statistically significant, or it could point to a deficiency of the model architecture chosen for the fully supervised classifier.}. Meanwhile, for $N_{\rm sig}=300$, we see that the gap between \textsc{R-Anode} and the fully supervised classifier widens, but the gap with the IAD widens even further. The IAD suffers relative to the fully supervised classifier due to limited training statistics, whereas \textsc{R-Anode} benefits from not being classifier-based and from its stronger inductive bias.

Next we examine the SIC values and total achieved significances at a fixed FPR of $10^{-3}$. These are shown in Fig.~\ref{fig:sic_and_significance} as a function of the number of injected signal events, for the different methods.
 We see from both plots that \textsc{R-Anode} (ideal) achieves a better sensitivity to signal at all signal  strengths, and allows us discover the signal at cross sections that are $\sim 25\%$ lower than the IAD.

\begin{figure}[t]
    \centering
    \includegraphics[width=0.43\textwidth]{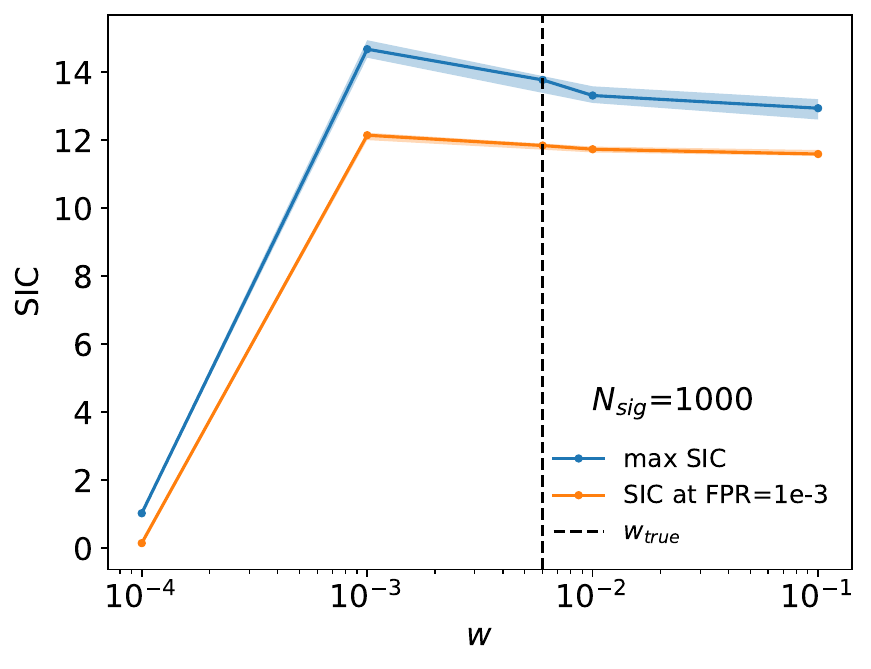}  
    \caption{For $N_{\rm sig}=1000$, we scan across different $w$-values, while holding it fixed during training. The significance improvement is robust for incorrect choices of $w$; however the performance drops significantly for $w\ll w_{\rm true}$.}
    \label{fig:w_scan}
\end{figure}

\subsection{Scanning fixed $w$}\label{sec:w_scan}

Next we study the more realistic case of scanning across different, fixed values of $w$ during training. We focus on the benchmark case of $N_{\rm sig}=1000$ for simplicity.

From Fig.~\ref{fig:w_scan}, we see that the performance curves and SIC values are actually quite robust to incorrect choices of $w$. We see that only at $w$-values an order of magnitude different than $w_{\rm true}$ is there some noticeable difference in the max SIC. At $w <w_{\rm true}$, the trained signal model could be overestimating regions in phase space where the signal density is higher than background. This might help it locate the region with higher signal density easier, which could explain why in Fig.~\ref{fig:w_scan}, we see a higher max SIC than $w_{\rm true}$. At $w \ll w_{\rm true}$, \textsc{R-Anode} loses the capability of modeling the signal, which leads to a sharp decline in performance. This could be possibly used to put a lower bound on $w$. Finally, for $w>w_{\rm true}$, the SIC declines slowly but does not completely go away. Here the method is learning to model the signal in addition to some amount of background. We expect as $w\to 1$, the \textsc{R-Anode} method smoothly interpolates to the original \textsc{Anode} method.

\subsection{Learnable $w$}\label{sec:learnable_w}

Finally, we explore what happens if we allow $w$ to be learned during training. For simplicity, we use the same optimizer for $w$ as the rest of the signal model, i.e.\ we just include $w$ in the list of learnable parameters during training. 

For each dataset -- as described in Section ~\ref{sec:dataset} -- we take $w$-values from 20 retrainings, and 10 lowest validation losses from each training. For learnable $w$, we find that some trainings converge to extremely low values of $w$. We label these trainings as having ``undetected signal" and choose to exclude them from the analysis. To do so, we form histograms of the 200 learned $w$-values as shown in Fig.~\ref{fig:w_full_histograms} of Appendix \ref{appendix:undetected}. These histograms show a clearly well-separated, bi-modal structure, with one mode corresponding to the ``undetected signal" case. We therefore devise a simple by-eye cut to remove this mode for a given dataset.

The remaining $w$-values are then averaged to produce a final output for the learned $w$ for a given dataset. (The corresponding $p_{\rm sig}$ distributions are also averaged as described above, to produce the anomaly score, etc.\ in the learned $w$ case.)

\begin{figure}[t]
    \centering
    \includegraphics[width=0.44\textwidth]{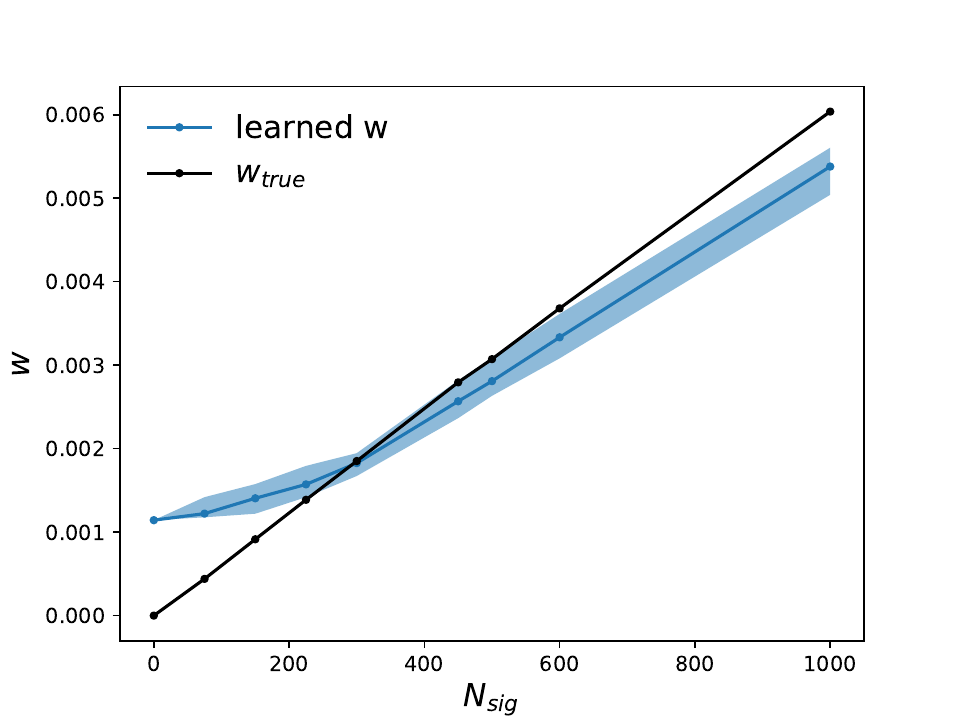}
    \caption{A comparison between learned $w$  and $w_{\rm true}$ for different $N_{\rm sig}$ values.}
    \label{fig:w_bounds}
    
\end{figure}

In Fig.~\ref{fig:w_bounds}, we show the final learned $w$ vs.\ true $w$, for a range of $N_{\rm sig}$ values. We also indicate the spread of outcomes across the 10 different datasets as a blue uncertainty band. We see that for the most part, there is a good agreement between the learned $w$ and true $w$ values. At larger $w$ there is a slight downward bias in the learned $w$, while at smaller $w$ there is a pronounced upward bias. Indeed,  for the no-signal fit, the learned $w$ values have an average value of $\approx 10^{-3}$. According to Fig.~\ref{fig:w_bounds}, our method with learnable $w$ is unable to detect $\lesssim 200$ signal events, for the given amount of background events. 

The undetectability of signal below some threshold may be an irreducible limitation of the \textsc{R-Anode} method with learnable $w$, or it could be a by-product of our decision to exclude the ``undetected signal" trainings and keep  trainings where a nonzero $w$ was returned. Ideally, one could devise a data-driven criterion that would correctly switch between the undetected and detected signal modes as $N_{\rm sig}$ decreases. Alternatively, one could stick to the current procedure and devise a process to quantify the lower threshold on signal strength in a data-driven way. For example, one could imagine applying \textsc{R-Anode} to synthetic background events sampled from the interpolated sideband model $p_{\rm bg}(x,m)$, which might enable us to measure the ``floor" for $w$. We save such investigations for future work.
  
As noted in Section.~\ref{sec:w_scan}, \textsc{R-Anode} as an anomaly detector is robust to incorrect choices of $w$. In Figs.~\ref{fig:performance_curves} and \ref{fig:sic_and_significance}, the \textsc{R-Anode} performance curves with learnable $w$ are shown in red. Although a slight performance drop relative to the idealized $w=w_{\rm true}$ case is seen for lower signal strengths, overall it is encouraging to see that the performance of the learnable $w$ case is actually extremely robust and nearly saturates the ideal \textsc{R-Anode} performance.

\subsection{Samples from the signal distribution}\label{sec:samples}

Using the IAD, one could make cuts on data with the anomaly score $R(x)$, to obtain the resulting signal samples. This requires a choice for the cut, which is not necessarily obvious from data. It also might leave too few signal events for interpretable distributions. Finally, the distributions one obtains would in general be biased, since the cut on $R(x)$ preferentially selects signal events that are more overdense relative to background. In \textsc{R-Anode} however we benefit from being able to directly (over)sample signal events in an (in principle) unbiased way.

\begin{figure*}
    \centering
    \includegraphics[width=0.45\textwidth,page=1]{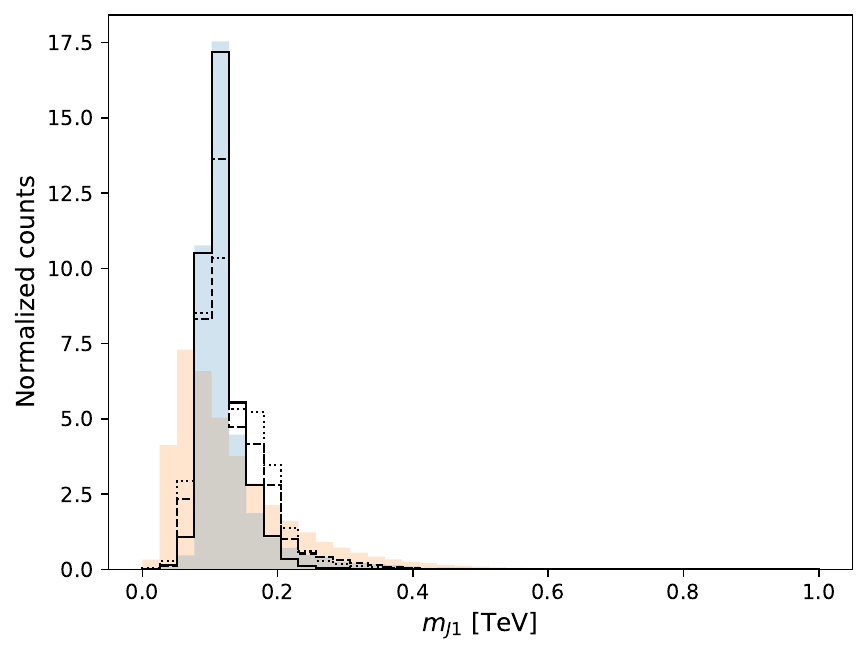}
    \includegraphics[width=0.45\textwidth,page=1]{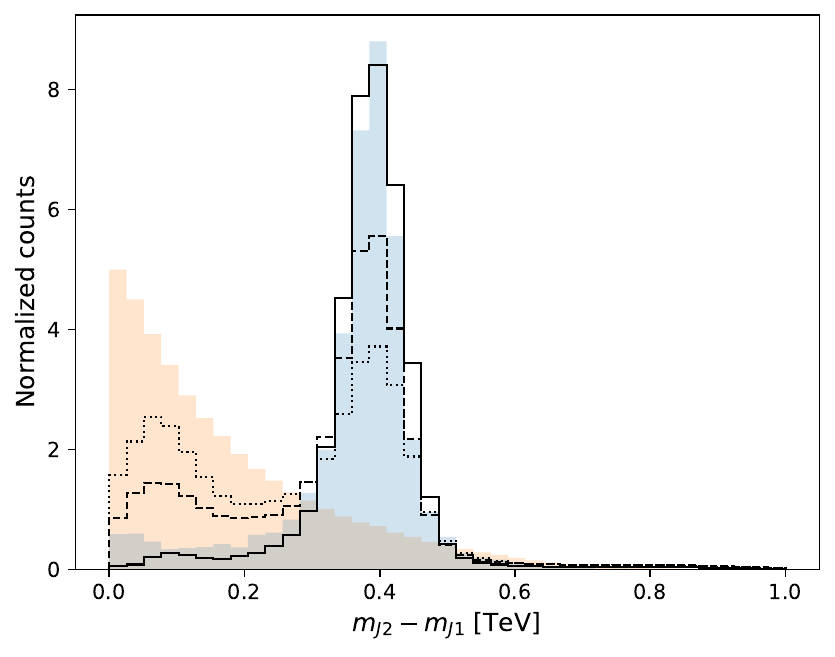}
    \includegraphics[width=0.45\textwidth,page=1]{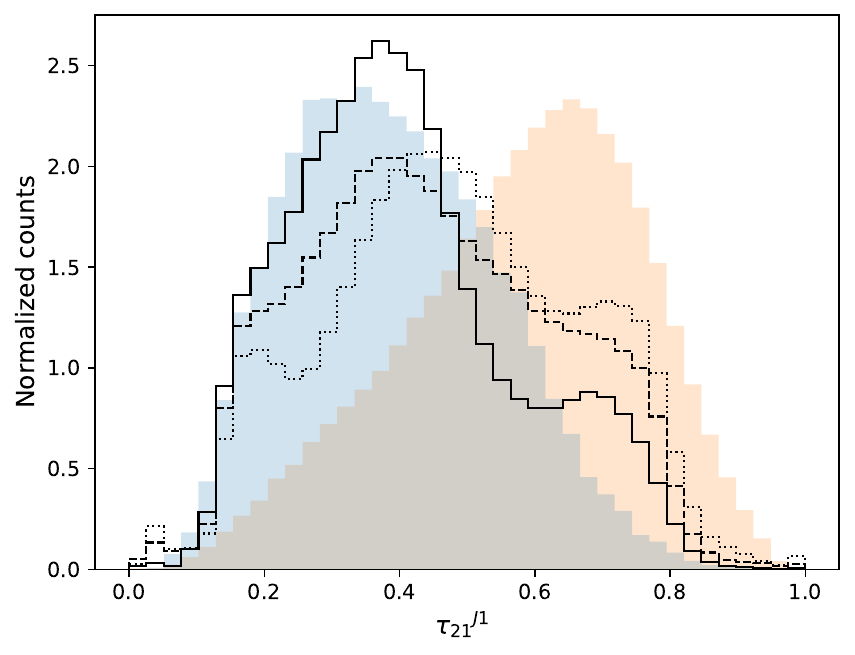}
    \includegraphics[width=0.45\textwidth,page=1]{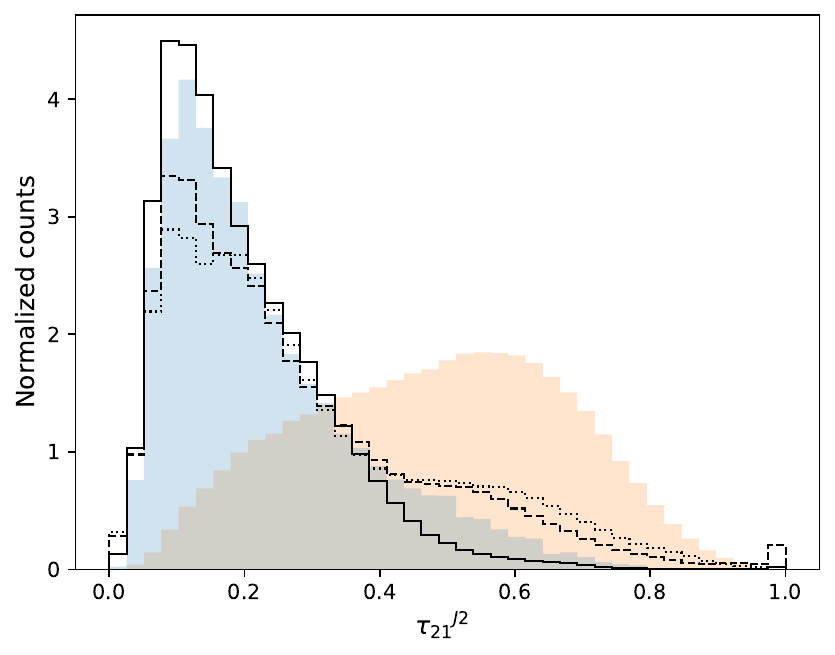}
    \includegraphics[width=0.45\textwidth,page=1]{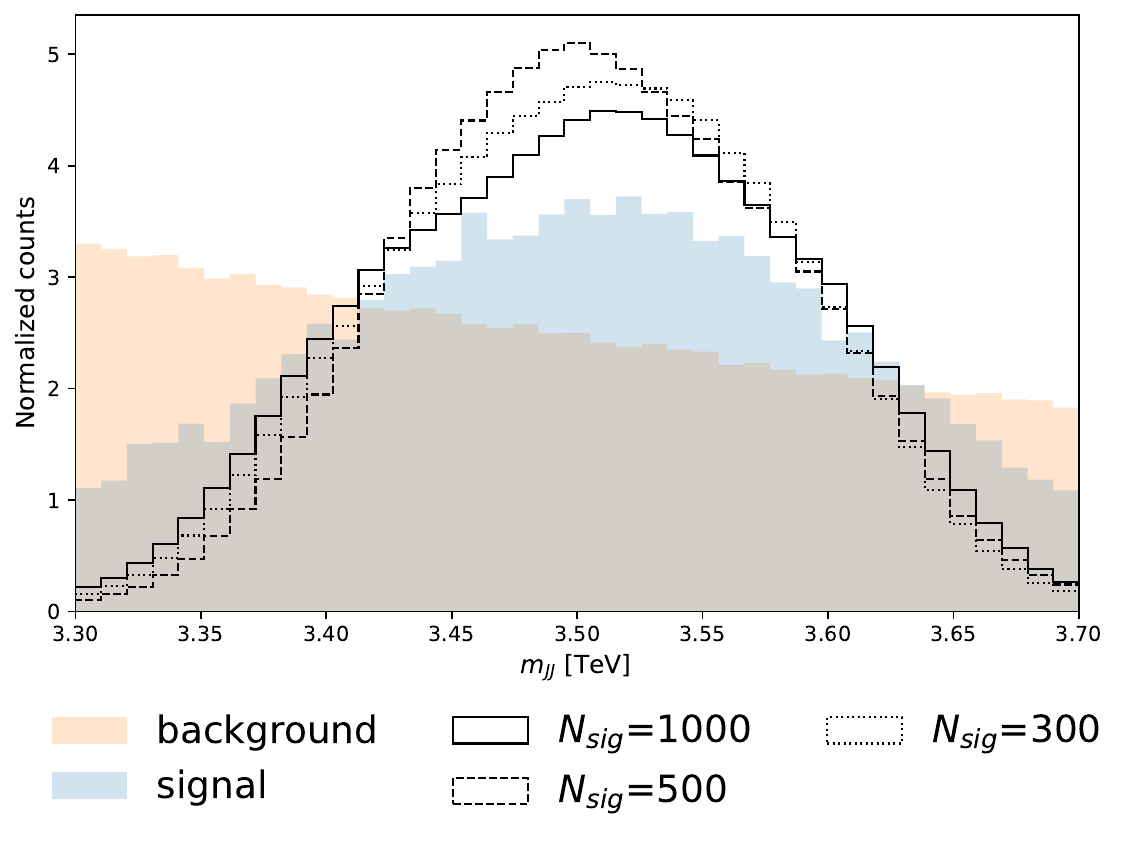}    \caption{Samples for \textsc{R-Anode} with a learnable $w$ during training, for $N_{\rm sig}=[1000,500,300]$. Solid histograms show the true background (signal) distributions in orange (blue), while black lines indicate the signal distributions learned for different injected signal amounts. Overall, the qualitative agreement between the learned and actual signal distributions is decent, although the samples clearly get worse for lower signal injections due insufficiency of training data.}
    \label{fig:learned_samples}

\end{figure*}

In Fig.~\ref{fig:learned_samples}, we show a comparison of these signal densities  obtained from \textsc{R-Anode} (learned) for different signal-injections.\footnote{We found that fixing to $ w_{\rm true}$ didn't improve these distributions much, hence we omit showing similar plots for \textsc{R-Anode} (ideal).} We see that the learned signal densities show good qualitative agreement with the true signal densities and allow us to get rough estimates on properties of the signal, like the invariant mass, subjet-mass, pronginess of each subjet, etc. There is a clear degradation in the quality of the learned signal densities at lower signal strengths. Clearly, an important future direction here will be to devise data-driven methods to place uncertainties on these learned distributions.

\section{Conclusions}\label{sec:conclusions}

\textsc{R-Anode} is a novel method for model-agnostic, data-driven resonant anomaly detection that sets a new standard for the state-of-the-art in both performance and interpretability. The key to \textsc{R-Anode} is to build in more inductive bias to the task of learning the optimal anomaly score. It builds on the previous approach of \textsc{Anode} \cite{anode}, but instead of fitting a density estimator for the data, it aims to isolate the small contribution from the unknown signal. By fitting a density estimator for just the signal component, holding fixed the background density learned from the sidebands, \textsc{R-Anode} is able to better approach the optimal anomaly detector, surpassing approaches based on classifiers and weak supervision.

By learning the signal density, \textsc{R-Anode} allows us to model the signal distributions directly from data, and this could be used to simultaneously discover and fully characterize the signal. 

One of the issues we have not addressed in this paper and which we save for a future study is background estimation. Combining \textsc{R-Anode} with a bump hunt would require us to examine the issue of mass sculpting in more detail, as was done in \cite{Hallin:2022eoq} for classifier-based approaches. Alternatively, perhaps the technique of ``direct background estimation" from \cite{anode} could be revisited, or the uncertainties on the learned $w$ could be properly calibrated somehow from data.

Another potential issue which we look forward to exploring further in future work is performance and robustness under the inclusion of additional, noisy high-level features \cite{cathodebdt,anode_bdt}.
Perhaps the BDT-based density estimator used in \cite{anode_bdt} could be beneficial for this task, while speeding up the density estimation as well. Finally, it will be very interesting to go beyond high-level features and implement the \textsc{R-Anode} method on the full phase space of jet constituents \cite{full_AD}, where classifier-based approaches are especially challenged.

\section*{acknowledgements}

We are grateful to Manuel Sommerhalder for assisting us with the scripts for training the background model in the SB. The work of RD and DS was supported by DOE grant DE-SC0010008. GK acknowledges support by the Deutsche Forschungsgemeinschaft under Germany’s Excellence Strategy – EXC 2121  Quantum Universe – 390833306. The authors acknowledge the Office of Advanced Research Computing (OARC) at Rutgers, The State University of New Jersey \url{https://it.rutgers.edu/oarc} for providing access to the Amarel cluster and associated research computing resources that have contributed to the results reported here.

\appendix

\section{Implementation details}\label{appendix:implementation}

The background model $p_{\rm bg}(x|m)$ is the same as \cite{anode,cathode}: A Masked Autoregressive Flow(MAF) with affine transformations \cite{maf}. It contains 15 MADE blocks, with each block consisting of one hidden layer of 128 nodes. We train the density estimator using \texttt{PyTorch}~\cite{pytorch} in the SB region for 100 epochs with \texttt{Adam} optimizer~\cite{Adam}, with a learning rate of $10^{-4}$, batch size 256, and batch normalization with a momentum of $1.0$. The base distribution chosen is unit normal. The SB data is split into a training-validation split of 50-50, and models with 10 lowest validation losses are selected. The probabilities obtained from these 10 models are averaged to obtain $p_{\rm bg}(x|m)$.\footnote{For simplicity we use the same background model, obtained from a $N_{\rm sig}=1000$ case, as $p_{\rm bg}(x|m)$ for all signal injections.} 

For $p_{\rm sig}(x,m)$ (fit to data in the SR only), we use RQS transformations \cite{rqs_paper} with 6 MADE blocks, with each block consisting of 2 hidden layers with 64 nodes, dropout 0.2, and batch-normalization is applied in between layers. The same model is also used for $p_{\rm data}(x|m)$ in the SR for \textsc{Anode}. The RQS-models for all cases are trained with a unit normal base distribution, and a learning rate of 0.0003 with the AdamW \cite{adamw} optimizer, with a batch size of 256 for 300 epochs. The train/val split and ensembling of $p_{\rm sig}$ and $p_{\rm data}$ were described in Sec.~\ref{sec:model}.

For the IAD and the fully supervised classifier, we train \texttt{HistGradientBoostingClassifier} with default hyperparameters for 200 epochs, similar to \cite{cathodebdt}.

Similar to \cite{cathode}, the features $x$ in the SB, are first shifted and scaled so that $x \in (0,1)$, logit transformed, and then standardized with mean $0$ and variance $1$. The same preprocessing parameters from the SB, were reused for the SR, for all the methods. For \textsc{R-Anode} and supervised, the mass is centered around zero, by subtracting 3.5 from all mass values.

\section{Trainings with undetected signal}
\label{appendix:undetected}

\begin{figure*}
    \centering
    \includegraphics[width=0.32\textwidth,page=1]{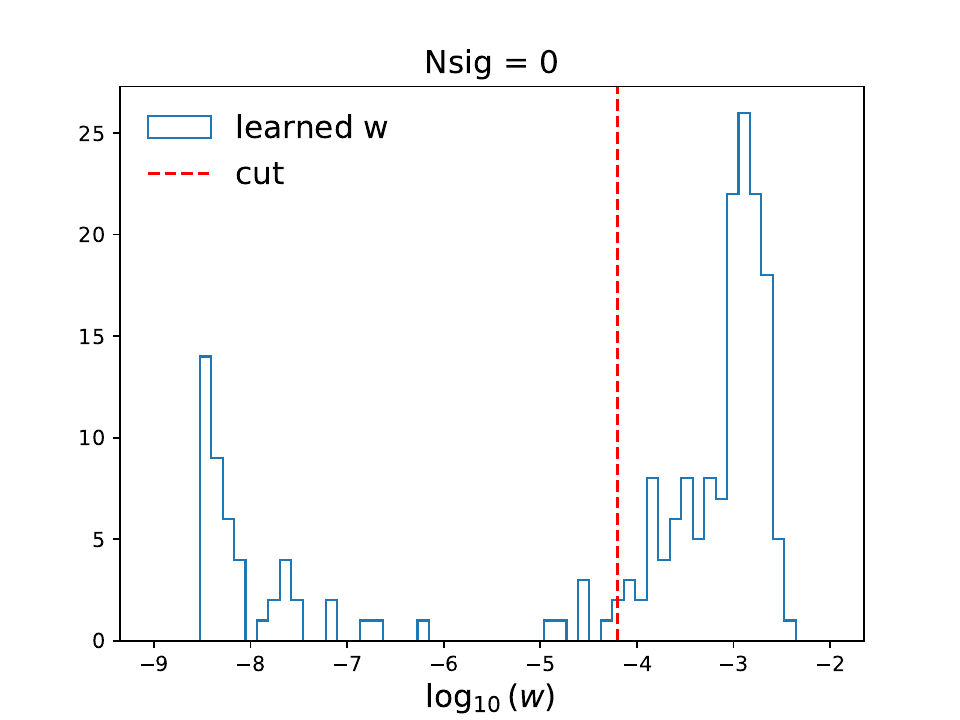}
    \includegraphics[width=0.32\textwidth,page=3]{figures/w_hist_500.pdf}
    \includegraphics[width=0.32\textwidth,page=1]{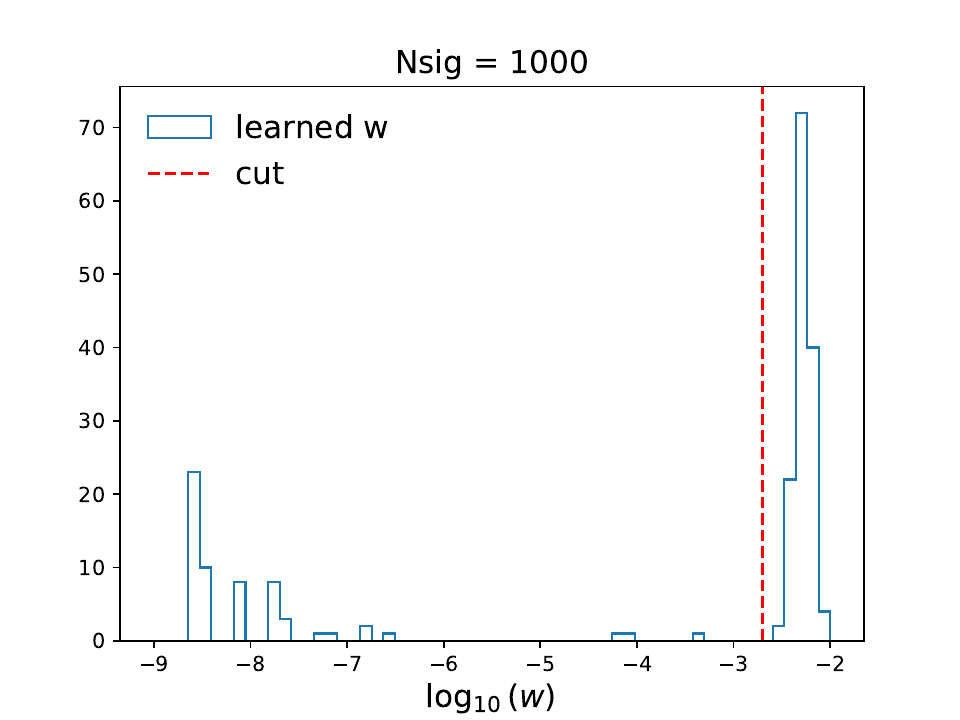}
    \caption{Histograms of all 200 learned $w$ values for three choices of $N_{\rm sig}$ (single dataset only). We see a clearly separated bi-modal structure, with ``undetected signal" trainings clustering at the low end (zero up to numerical precision). The red dashed lines are our thresholds for  trainings with undetected signal; $w$ values below these thresholds are rejected. The thresholds are decided dataset-by-dataset on the basis of a by-eye inspection.}
    \label{fig:w_full_histograms}
    
\end{figure*}

As mentioned in Section~\ref{sec:learnable_w}, Fig.~\ref{fig:w_full_histograms} shows the full histograms of learned $w$ values, for different $N_{\rm sig}$ values (single dataset only). The ``undetected signal" trainings where $w$ converged to zero (up to machine precision) are clearly seen in the bi-modal histograms. The cuts shown in Fig.~\ref{fig:w_full_histograms} are based on by-eye estimates, performed for each dataset, and all the results for \textsc{R-Anode} (learned) are based on models corresponding to $w$-values surviving above these cuts.

\bibliographystyle{apsrev4-1}

\bibliography{references}

\begin{thebibliography}{51}%
\makeatletter
\providecommand \@ifxundefined [1]{%
 \@ifx{#1\undefined}
}%
\providecommand \@ifnum [1]{%
 \ifnum #1\expandafter \@firstoftwo
 \else \expandafter \@secondoftwo
 \fi
}%
\providecommand \@ifx [1]{%
 \ifx #1\expandafter \@firstoftwo
 \else \expandafter \@secondoftwo
 \fi
}%
\providecommand \natexlab [1]{#1}%
\providecommand \enquote  [1]{``#1''}%
\providecommand \bibnamefont  [1]{#1}%
\providecommand \bibfnamefont [1]{#1}%
\providecommand \citenamefont [1]{#1}%
\providecommand \href@noop [0]{\@secondoftwo}%
\providecommand \href [0]{\begingroup \@sanitize@url \@href}%
\providecommand \@href[1]{\@@startlink{#1}\@@href}%
\providecommand \@@href[1]{\endgroup#1\@@endlink}%
\providecommand \@sanitize@url [0]{\catcode `\\12\catcode `\$12\catcode `\&12\catcode `\#12\catcode `\^12\catcode `\_12\catcode `\%12\relax}%
\providecommand \@@startlink[1]{}%
\providecommand \@@endlink[0]{}%
\providecommand \url  [0]{\begingroup\@sanitize@url \@url }%
\providecommand \@url [1]{\endgroup\@href {#1}{\urlprefix }}%
\providecommand \urlprefix  [0]{URL }%
\providecommand \Eprint [0]{\href }%
\providecommand \doibase [0]{http://dx.doi.org/}%
\providecommand \selectlanguage [0]{\@gobble}%
\providecommand \bibinfo  [0]{\@secondoftwo}%
\providecommand \bibfield  [0]{\@secondoftwo}%
\providecommand \translation [1]{[#1]}%
\providecommand \BibitemOpen [0]{}%
\providecommand \bibitemStop [0]{}%
\providecommand \bibitemNoStop [0]{.\EOS\space}%
\providecommand \EOS [0]{\spacefactor3000\relax}%
\providecommand \BibitemShut  [1]{\csname bibitem#1\endcsname}%
\let\auto@bib@innerbib\@empty
\bibitem [{\citenamefont {{ATLAS Collaboration}}(2023{\natexlab{a}})}]{atlasexoticstwiki}%
  \BibitemOpen
  \bibfield  {author} {\bibinfo {author} {\bibnamefont {{ATLAS Collaboration}}},\ }\href@noop {} {\enquote {\bibinfo {title} {{Exotic Physics Searches}},}\ } (\bibinfo {year} {2023}{\natexlab{a}}),\ \bibinfo {note} {\url{https://twiki.cern.ch/twiki/bin/view/AtlasPublic/ExoticsPublicResults}}\BibitemShut {NoStop}%
\bibitem [{\citenamefont {{ATLAS Collaboration}}(2023{\natexlab{b}})}]{atlassusytwiki}%
  \BibitemOpen
  \bibfield  {author} {\bibinfo {author} {\bibnamefont {{ATLAS Collaboration}}},\ }\href@noop {} {\enquote {\bibinfo {title} {{Supersymmetry searches}},}\ } (\bibinfo {year} {2023}{\natexlab{b}}),\ \bibinfo {note} {\url{https://twiki.cern.ch/twiki/bin/view/AtlasPublic/SupersymmetryPublicResults}}\BibitemShut {NoStop}%
\bibitem [{\citenamefont {{ATLAS Collaboration}}(2023{\natexlab{c}})}]{atlashdbspublictwiki}%
  \BibitemOpen
  \bibfield  {author} {\bibinfo {author} {\bibnamefont {{ATLAS Collaboration}}},\ }\href@noop {} {\enquote {\bibinfo {title} {{Higgs and Diboson Searches}},}\ } (\bibinfo {year} {2023}{\natexlab{c}}),\ \bibinfo {note} {\url{https://twiki.cern.ch/twiki/bin/view/AtlasPublic/HDBSPublicResults}}\BibitemShut {NoStop}%
\bibitem [{\citenamefont {{CMS Collaboration}}(2023{\natexlab{a}})}]{cmsexoticstwiki}%
  \BibitemOpen
  \bibfield  {author} {\bibinfo {author} {\bibnamefont {{CMS Collaboration}}},\ }\href@noop {} {\enquote {\bibinfo {title} {{CMS Exotica Public Physics Results}},}\ } (\bibinfo {year} {2023}{\natexlab{a}}),\ \bibinfo {note} {\url{https://twiki.cern.ch/twiki/bin/view/CMSPublic/PhysicsResultsEXO}}\BibitemShut {NoStop}%
\bibitem [{\citenamefont {{CMS Collaboration}}(2023{\natexlab{b}})}]{cmssusytwiki}%
  \BibitemOpen
  \bibfield  {author} {\bibinfo {author} {\bibnamefont {{CMS Collaboration}}},\ }\href@noop {} {\enquote {\bibinfo {title} {{CMS Supersymmetry Physics Results}},}\ } (\bibinfo {year} {2023}{\natexlab{b}}),\ \bibinfo {note} {\url{https://twiki.cern.ch/twiki/bin/view/CMSPublic/PhysicsResultsSUS}}\BibitemShut {NoStop}%
\bibitem [{\citenamefont {{CMS Collaboration}}(2023{\natexlab{c}})}]{cmsb2gtwiki}%
  \BibitemOpen
  \bibfield  {author} {\bibinfo {author} {\bibnamefont {{CMS Collaboration}}},\ }\href@noop {} {\enquote {\bibinfo {title} {{CMS Beyond-two-generations (B2G) Public Physics Results}},}\ } (\bibinfo {year} {2023}{\natexlab{c}}),\ \bibinfo {note} {\url{https://twiki.cern.ch/twiki/bin/view/CMSPublic/PhysicsResultsB2G}}\BibitemShut {NoStop}%
\bibitem [{\citenamefont {{LHCb Collaboration}}(2023)}]{lhcbtwiki}%
  \BibitemOpen
  \bibfield  {author} {\bibinfo {author} {\bibnamefont {{LHCb Collaboration}}},\ }\href@noop {} {\enquote {\bibinfo {title} {{Publications of the QCD, Electroweak and Exotica Working Group}},}\ } (\bibinfo {year} {2023}),\ \bibinfo {note} {\url{http://lhcbproject.web.cern.ch/lhcbproject/Publications/LHCbProjectPublic/Summary_QEE.html}}\BibitemShut {NoStop}%
\bibitem [{\citenamefont {Kasieczka}\ \emph {et~al.}(2021{\natexlab{a}})\citenamefont {Kasieczka}, \citenamefont {Nachman}, \citenamefont {Shih} \emph {et~al.}}]{Kasieczka:2021xcg}%
  \BibitemOpen
  \bibfield  {author} {\bibinfo {author} {\bibfnamefont {G.}~\bibnamefont {Kasieczka}}, \bibinfo {author} {\bibfnamefont {B.}~\bibnamefont {Nachman}}, \bibinfo {author} {\bibfnamefont {D.}~\bibnamefont {Shih}},  \emph {et~al.},\ }\href {\doibase 10.1088/1361-6633/ac36b9} {\bibfield  {journal} {\bibinfo  {journal} {Rept. Prog. Phys.}\ }\textbf {\bibinfo {volume} {84}},\ \bibinfo {pages} {124201} (\bibinfo {year} {2021}{\natexlab{a}})},\ \Eprint {http://arxiv.org/abs/2101.08320} {arXiv:2101.08320 [hep-ph]} \BibitemShut {NoStop}%
\bibitem [{\citenamefont {Aarrestad}\ \emph {et~al.}(2021)\citenamefont {Aarrestad} \emph {et~al.}}]{Aarrestad:2021oeb}%
  \BibitemOpen
  \bibfield  {author} {\bibinfo {author} {\bibfnamefont {T.}~\bibnamefont {Aarrestad}} \emph {et~al.},\ }\href@noop {} {\  (\bibinfo {year} {2021})},\ \Eprint {http://arxiv.org/abs/2105.14027} {arXiv:2105.14027 [hep-ph]} \BibitemShut {NoStop}%
\bibitem [{\citenamefont {Karagiorgi}\ \emph {et~al.}(2022)\citenamefont {Karagiorgi}, \citenamefont {Kasieczka}, \citenamefont {Kravitz}, \citenamefont {Nachman},\ and\ \citenamefont {Shih}}]{Karagiorgi:2022qnh}%
  \BibitemOpen
  \bibfield  {author} {\bibinfo {author} {\bibfnamefont {G.}~\bibnamefont {Karagiorgi}}, \bibinfo {author} {\bibfnamefont {G.}~\bibnamefont {Kasieczka}}, \bibinfo {author} {\bibfnamefont {S.}~\bibnamefont {Kravitz}}, \bibinfo {author} {\bibfnamefont {B.}~\bibnamefont {Nachman}}, \ and\ \bibinfo {author} {\bibfnamefont {D.}~\bibnamefont {Shih}},\ }\href {\doibase 10.1038/s42254-022-00455-1} {\bibfield  {journal} {\bibinfo  {journal} {Nature Rev. Phys.}\ }\textbf {\bibinfo {volume} {4}},\ \bibinfo {pages} {399} (\bibinfo {year} {2022})}\BibitemShut {NoStop}%
\bibitem [{\citenamefont {Collins}\ \emph {et~al.}(2018)\citenamefont {Collins}, \citenamefont {Howe},\ and\ \citenamefont {Nachman}}]{Collins:2018epr}%
  \BibitemOpen
  \bibfield  {author} {\bibinfo {author} {\bibfnamefont {J.~H.}\ \bibnamefont {Collins}}, \bibinfo {author} {\bibfnamefont {K.}~\bibnamefont {Howe}}, \ and\ \bibinfo {author} {\bibfnamefont {B.}~\bibnamefont {Nachman}},\ }\href {\doibase 10.1103/PhysRevLett.121.241803} {\bibfield  {journal} {\bibinfo  {journal} {Phys. Rev. Lett.}\ }\textbf {\bibinfo {volume} {121}},\ \bibinfo {pages} {241803} (\bibinfo {year} {2018})},\ \Eprint {http://arxiv.org/abs/1805.02664} {arXiv:1805.02664 [hep-ph]} \BibitemShut {NoStop}%
\bibitem [{\citenamefont {Collins}\ \emph {et~al.}(2019)\citenamefont {Collins}, \citenamefont {Howe},\ and\ \citenamefont {Nachman}}]{Collins:2019jip}%
  \BibitemOpen
  \bibfield  {author} {\bibinfo {author} {\bibfnamefont {J.~H.}\ \bibnamefont {Collins}}, \bibinfo {author} {\bibfnamefont {K.}~\bibnamefont {Howe}}, \ and\ \bibinfo {author} {\bibfnamefont {B.}~\bibnamefont {Nachman}},\ }\href {\doibase 10.1103/PhysRevD.99.014038} {\bibfield  {journal} {\bibinfo  {journal} {Phys. Rev.}\ }\textbf {\bibinfo {volume} {D99}},\ \bibinfo {pages} {014038} (\bibinfo {year} {2019})},\ \Eprint {http://arxiv.org/abs/1902.02634} {arXiv:1902.02634 [hep-ph]} \BibitemShut {NoStop}%
\bibitem [{\citenamefont {Nachman}\ and\ \citenamefont {Shih}(2020)}]{anode}%
  \BibitemOpen
  \bibfield  {author} {\bibinfo {author} {\bibfnamefont {B.}~\bibnamefont {Nachman}}\ and\ \bibinfo {author} {\bibfnamefont {D.}~\bibnamefont {Shih}},\ }\href {\doibase 10.1103/physrevd.101.075042} {\bibfield  {journal} {\bibinfo  {journal} {Physical Review D}\ }\textbf {\bibinfo {volume} {101}} (\bibinfo {year} {2020}),\ 10.1103/physrevd.101.075042}\BibitemShut {NoStop}%
\bibitem [{\citenamefont {Andreassen}\ \emph {et~al.}(2020)\citenamefont {Andreassen}, \citenamefont {Nachman},\ and\ \citenamefont {Shih}}]{Andreassen:2020nkr}%
  \BibitemOpen
  \bibfield  {author} {\bibinfo {author} {\bibfnamefont {A.}~\bibnamefont {Andreassen}}, \bibinfo {author} {\bibfnamefont {B.}~\bibnamefont {Nachman}}, \ and\ \bibinfo {author} {\bibfnamefont {D.}~\bibnamefont {Shih}},\ }\href {\doibase 10.1103/PhysRevD.101.095004} {\bibfield  {journal} {\bibinfo  {journal} {Phys. Rev. D}\ }\textbf {\bibinfo {volume} {101}},\ \bibinfo {pages} {095004} (\bibinfo {year} {2020})},\ \Eprint {http://arxiv.org/abs/2001.05001} {arXiv:2001.05001 [hep-ph]} \BibitemShut {NoStop}%
\bibitem [{\citenamefont {Stein}\ \emph {et~al.}(2020)\citenamefont {Stein}, \citenamefont {Seljak},\ and\ \citenamefont {Dai}}]{Stein:2020rou}%
  \BibitemOpen
  \bibfield  {author} {\bibinfo {author} {\bibfnamefont {G.}~\bibnamefont {Stein}}, \bibinfo {author} {\bibfnamefont {U.}~\bibnamefont {Seljak}}, \ and\ \bibinfo {author} {\bibfnamefont {B.}~\bibnamefont {Dai}},\ }in\ \href@noop {} {\emph {\bibinfo {booktitle} {{34th Conference on Neural Information Processing Systems}}}}\ (\bibinfo {year} {2020})\ \Eprint {http://arxiv.org/abs/2012.11638} {arXiv:2012.11638 [cs.LG]} \BibitemShut {NoStop}%
\bibitem [{\citenamefont {Amram}\ and\ \citenamefont {Suarez}(2020)}]{Amram:2020ykb}%
  \BibitemOpen
  \bibfield  {author} {\bibinfo {author} {\bibfnamefont {O.}~\bibnamefont {Amram}}\ and\ \bibinfo {author} {\bibfnamefont {C.~M.}\ \bibnamefont {Suarez}},\ }\href {\doibase 10.1007/JHEP01(2021)153} {\  (\bibinfo {year} {2020}),\ 10.1007/JHEP01(2021)153},\ \Eprint {http://arxiv.org/abs/2002.12376} {arXiv:2002.12376 [hep-ph]} \BibitemShut {NoStop}%
\bibitem [{\citenamefont {Hallin}\ \emph {et~al.}(2022)\citenamefont {Hallin}, \citenamefont {Isaacson}, \citenamefont {Kasieczka}, \citenamefont {Krause}, \citenamefont {Nachman}, \citenamefont {Quadfasel}, \citenamefont {Schlaffer}, \citenamefont {Shih},\ and\ \citenamefont {Sommerhalder}}]{cathode}%
  \BibitemOpen
  \bibfield  {author} {\bibinfo {author} {\bibfnamefont {A.}~\bibnamefont {Hallin}}, \bibinfo {author} {\bibfnamefont {J.}~\bibnamefont {Isaacson}}, \bibinfo {author} {\bibfnamefont {G.}~\bibnamefont {Kasieczka}}, \bibinfo {author} {\bibfnamefont {C.}~\bibnamefont {Krause}}, \bibinfo {author} {\bibfnamefont {B.}~\bibnamefont {Nachman}}, \bibinfo {author} {\bibfnamefont {T.}~\bibnamefont {Quadfasel}}, \bibinfo {author} {\bibfnamefont {M.}~\bibnamefont {Schlaffer}}, \bibinfo {author} {\bibfnamefont {D.}~\bibnamefont {Shih}}, \ and\ \bibinfo {author} {\bibfnamefont {M.}~\bibnamefont {Sommerhalder}},\ }\href {\doibase 10.1103/physrevd.106.055006} {\bibfield  {journal} {\bibinfo  {journal} {Physical Review D}\ }\textbf {\bibinfo {volume} {106}} (\bibinfo {year} {2022}),\ 10.1103/physrevd.106.055006}\BibitemShut {NoStop}%
\bibitem [{\citenamefont {Collins}\ \emph {et~al.}(2021)\citenamefont {Collins}, \citenamefont {Mart\'\i{}n-Ramiro}, \citenamefont {Nachman},\ and\ \citenamefont {Shih}}]{Collins:2021nxn}%
  \BibitemOpen
  \bibfield  {author} {\bibinfo {author} {\bibfnamefont {J.~H.}\ \bibnamefont {Collins}}, \bibinfo {author} {\bibfnamefont {P.}~\bibnamefont {Mart\'\i{}n-Ramiro}}, \bibinfo {author} {\bibfnamefont {B.}~\bibnamefont {Nachman}}, \ and\ \bibinfo {author} {\bibfnamefont {D.}~\bibnamefont {Shih}},\ }\href@noop {} {\  (\bibinfo {year} {2021})},\ \Eprint {http://arxiv.org/abs/2104.02092} {arXiv:2104.02092 [hep-ph]} \BibitemShut {NoStop}%
\bibitem [{\citenamefont {Benkendorfer}\ \emph {et~al.}(2020)\citenamefont {Benkendorfer}, \citenamefont {Pottier},\ and\ \citenamefont {Nachman}}]{1815227}%
  \BibitemOpen
  \bibfield  {author} {\bibinfo {author} {\bibfnamefont {K.}~\bibnamefont {Benkendorfer}}, \bibinfo {author} {\bibfnamefont {L.~L.}\ \bibnamefont {Pottier}}, \ and\ \bibinfo {author} {\bibfnamefont {B.}~\bibnamefont {Nachman}},\ }\href@noop {} {\  (\bibinfo {year} {2020})},\ \Eprint {http://arxiv.org/abs/2009.02205} {arXiv:2009.02205 [hep-ph]} \BibitemShut {NoStop}%
\bibitem [{\citenamefont {Kasieczka}\ \emph {et~al.}(2021{\natexlab{b}})\citenamefont {Kasieczka}, \citenamefont {Nachman},\ and\ \citenamefont {Shih}}]{Kasieczka:2021tew}%
  \BibitemOpen
  \bibfield  {author} {\bibinfo {author} {\bibfnamefont {G.}~\bibnamefont {Kasieczka}}, \bibinfo {author} {\bibfnamefont {B.}~\bibnamefont {Nachman}}, \ and\ \bibinfo {author} {\bibfnamefont {D.}~\bibnamefont {Shih}}\ }(\bibinfo {year} {2021})\ \Eprint {http://arxiv.org/abs/2107.02821} {arXiv:2107.02821 [stat.ML]} \BibitemShut {NoStop}%
\bibitem [{\citenamefont {Hallin}\ \emph {et~al.}(2023)\citenamefont {Hallin}, \citenamefont {Kasieczka}, \citenamefont {Quadfasel}, \citenamefont {Shih},\ and\ \citenamefont {Sommerhalder}}]{Hallin:2022eoq}%
  \BibitemOpen
  \bibfield  {author} {\bibinfo {author} {\bibfnamefont {A.}~\bibnamefont {Hallin}}, \bibinfo {author} {\bibfnamefont {G.}~\bibnamefont {Kasieczka}}, \bibinfo {author} {\bibfnamefont {T.}~\bibnamefont {Quadfasel}}, \bibinfo {author} {\bibfnamefont {D.}~\bibnamefont {Shih}}, \ and\ \bibinfo {author} {\bibfnamefont {M.}~\bibnamefont {Sommerhalder}},\ }\href {\doibase 10.1103/PhysRevD.107.114012} {\bibfield  {journal} {\bibinfo  {journal} {Phys. Rev. D}\ }\textbf {\bibinfo {volume} {107}},\ \bibinfo {pages} {114012} (\bibinfo {year} {2023})},\ \Eprint {http://arxiv.org/abs/2210.14924} {arXiv:2210.14924 [hep-ph]} \BibitemShut {NoStop}%
\bibitem [{\citenamefont {Chen}\ \emph {et~al.}(2023)\citenamefont {Chen}, \citenamefont {Nachman},\ and\ \citenamefont {Sala}}]{Chen:2022suv}%
  \BibitemOpen
  \bibfield  {author} {\bibinfo {author} {\bibfnamefont {M.~F.}\ \bibnamefont {Chen}}, \bibinfo {author} {\bibfnamefont {B.}~\bibnamefont {Nachman}}, \ and\ \bibinfo {author} {\bibfnamefont {F.}~\bibnamefont {Sala}},\ }\href {\doibase 10.1007/JHEP07(2023)188} {\bibfield  {journal} {\bibinfo  {journal} {JHEP}\ }\textbf {\bibinfo {volume} {07}},\ \bibinfo {pages} {188} (\bibinfo {year} {2023})},\ \Eprint {http://arxiv.org/abs/2212.10579} {arXiv:2212.10579 [hep-ph]} \BibitemShut {NoStop}%
\bibitem [{\citenamefont {Kamenik}\ and\ \citenamefont {Szewc}(2023)}]{Kamenik:2022qxs}%
  \BibitemOpen
  \bibfield  {author} {\bibinfo {author} {\bibfnamefont {J.~F.}\ \bibnamefont {Kamenik}}\ and\ \bibinfo {author} {\bibfnamefont {M.}~\bibnamefont {Szewc}},\ }\href {\doibase 10.1016/j.physletb.2023.137836} {\bibfield  {journal} {\bibinfo  {journal} {Phys. Lett. B}\ }\textbf {\bibinfo {volume} {840}},\ \bibinfo {pages} {137836} (\bibinfo {year} {2023})},\ \Eprint {http://arxiv.org/abs/2210.02226} {arXiv:2210.02226 [hep-ph]} \BibitemShut {NoStop}%
\bibitem [{\citenamefont {Sengupta}\ \emph {et~al.}(2023)\citenamefont {Sengupta}, \citenamefont {Klein}, \citenamefont {Raine},\ and\ \citenamefont {Golling}}]{Sengupta:2023xqy}%
  \BibitemOpen
  \bibfield  {author} {\bibinfo {author} {\bibfnamefont {D.}~\bibnamefont {Sengupta}}, \bibinfo {author} {\bibfnamefont {S.}~\bibnamefont {Klein}}, \bibinfo {author} {\bibfnamefont {J.~A.}\ \bibnamefont {Raine}}, \ and\ \bibinfo {author} {\bibfnamefont {T.}~\bibnamefont {Golling}},\ }\href@noop {} {\  (\bibinfo {year} {2023})},\ \Eprint {http://arxiv.org/abs/2305.04646} {arXiv:2305.04646 [hep-ph]} \BibitemShut {NoStop}%
\bibitem [{\citenamefont {Raine}\ \emph {et~al.}(2023)\citenamefont {Raine}, \citenamefont {Klein}, \citenamefont {Sengupta},\ and\ \citenamefont {Golling}}]{Raine:2022hht}%
  \BibitemOpen
  \bibfield  {author} {\bibinfo {author} {\bibfnamefont {J.~A.}\ \bibnamefont {Raine}}, \bibinfo {author} {\bibfnamefont {S.}~\bibnamefont {Klein}}, \bibinfo {author} {\bibfnamefont {D.}~\bibnamefont {Sengupta}}, \ and\ \bibinfo {author} {\bibfnamefont {T.}~\bibnamefont {Golling}},\ }\href {\doibase 10.3389/fdata.2023.899345} {\bibfield  {journal} {\bibinfo  {journal} {Front. Big Data}\ }\textbf {\bibinfo {volume} {6}},\ \bibinfo {pages} {899345} (\bibinfo {year} {2023})},\ \Eprint {http://arxiv.org/abs/2203.09470} {arXiv:2203.09470 [hep-ph]} \BibitemShut {NoStop}%
\bibitem [{\citenamefont {Golling}\ \emph {et~al.}(2023{\natexlab{a}})\citenamefont {Golling}, \citenamefont {Kasieczka}, \citenamefont {Krause}, \citenamefont {Mastandrea}, \citenamefont {Nachman}, \citenamefont {Raine}, \citenamefont {Sengupta}, \citenamefont {Shih},\ and\ \citenamefont {Sommerhalder}}]{Golling:2023yjq}%
  \BibitemOpen
  \bibfield  {author} {\bibinfo {author} {\bibfnamefont {T.}~\bibnamefont {Golling}}, \bibinfo {author} {\bibfnamefont {G.}~\bibnamefont {Kasieczka}}, \bibinfo {author} {\bibfnamefont {C.}~\bibnamefont {Krause}}, \bibinfo {author} {\bibfnamefont {R.}~\bibnamefont {Mastandrea}}, \bibinfo {author} {\bibfnamefont {B.}~\bibnamefont {Nachman}}, \bibinfo {author} {\bibfnamefont {J.~A.}\ \bibnamefont {Raine}}, \bibinfo {author} {\bibfnamefont {D.}~\bibnamefont {Sengupta}}, \bibinfo {author} {\bibfnamefont {D.}~\bibnamefont {Shih}}, \ and\ \bibinfo {author} {\bibfnamefont {M.}~\bibnamefont {Sommerhalder}},\ }\href@noop {} {\  (\bibinfo {year} {2023}{\natexlab{a}})},\ \Eprint {http://arxiv.org/abs/2307.11157} {arXiv:2307.11157 [hep-ph]} \BibitemShut {NoStop}%
\bibitem [{\citenamefont {Golling}\ \emph {et~al.}(2023{\natexlab{b}})\citenamefont {Golling}, \citenamefont {Klein}, \citenamefont {Mastandrea},\ and\ \citenamefont {Nachman}}]{feta}%
  \BibitemOpen
  \bibfield  {author} {\bibinfo {author} {\bibfnamefont {T.}~\bibnamefont {Golling}}, \bibinfo {author} {\bibfnamefont {S.}~\bibnamefont {Klein}}, \bibinfo {author} {\bibfnamefont {R.}~\bibnamefont {Mastandrea}}, \ and\ \bibinfo {author} {\bibfnamefont {B.}~\bibnamefont {Nachman}},\ }\href {\doibase 10.1103/physrevd.107.096025} {\bibfield  {journal} {\bibinfo  {journal} {Physical Review D}\ }\textbf {\bibinfo {volume} {107}} (\bibinfo {year} {2023}{\natexlab{b}}),\ 10.1103/physrevd.107.096025}\BibitemShut {NoStop}%
\bibitem [{\citenamefont {Bickendorf}\ \emph {et~al.}(2023)\citenamefont {Bickendorf}, \citenamefont {Drees}, \citenamefont {Kasieczka}, \citenamefont {Krause},\ and\ \citenamefont {Shih}}]{bickendorf2023combining}%
  \BibitemOpen
  \bibfield  {author} {\bibinfo {author} {\bibfnamefont {G.}~\bibnamefont {Bickendorf}}, \bibinfo {author} {\bibfnamefont {M.}~\bibnamefont {Drees}}, \bibinfo {author} {\bibfnamefont {G.}~\bibnamefont {Kasieczka}}, \bibinfo {author} {\bibfnamefont {C.}~\bibnamefont {Krause}}, \ and\ \bibinfo {author} {\bibfnamefont {D.}~\bibnamefont {Shih}},\ }\href@noop {} {\enquote {\bibinfo {title} {{Combining Resonant and Tail-based Anomaly Detection}},}\ } (\bibinfo {year} {2023}),\ \Eprint {http://arxiv.org/abs/2309.12918} {arXiv:2309.12918 [hep-ph]} \BibitemShut {NoStop}%
\bibitem [{\citenamefont {Finke}\ \emph {et~al.}(2023)\citenamefont {Finke}, \citenamefont {Hein}, \citenamefont {Kasieczka}, \citenamefont {Krämer}, \citenamefont {Mück}, \citenamefont {Prangchaikul}, \citenamefont {Quadfasel}, \citenamefont {Shih},\ and\ \citenamefont {Sommerhalder}}]{cathodebdt}%
  \BibitemOpen
  \bibfield  {author} {\bibinfo {author} {\bibfnamefont {T.}~\bibnamefont {Finke}}, \bibinfo {author} {\bibfnamefont {M.}~\bibnamefont {Hein}}, \bibinfo {author} {\bibfnamefont {G.}~\bibnamefont {Kasieczka}}, \bibinfo {author} {\bibfnamefont {M.}~\bibnamefont {Krämer}}, \bibinfo {author} {\bibfnamefont {A.}~\bibnamefont {Mück}}, \bibinfo {author} {\bibfnamefont {P.}~\bibnamefont {Prangchaikul}}, \bibinfo {author} {\bibfnamefont {T.}~\bibnamefont {Quadfasel}}, \bibinfo {author} {\bibfnamefont {D.}~\bibnamefont {Shih}}, \ and\ \bibinfo {author} {\bibfnamefont {M.}~\bibnamefont {Sommerhalder}},\ }\href@noop {} {\enquote {\bibinfo {title} {Back to the roots: Tree-based algorithms for weakly supervised anomaly detection},}\ } (\bibinfo {year} {2023}),\ \Eprint {http://arxiv.org/abs/2309.13111} {arXiv:2309.13111 [hep-ph]} \BibitemShut {NoStop}%
\bibitem [{\citenamefont {Freytsis}\ \emph {et~al.}(2023)\citenamefont {Freytsis}, \citenamefont {Perelstein},\ and\ \citenamefont {San}}]{anode_bdt}%
  \BibitemOpen
  \bibfield  {author} {\bibinfo {author} {\bibfnamefont {M.}~\bibnamefont {Freytsis}}, \bibinfo {author} {\bibfnamefont {M.}~\bibnamefont {Perelstein}}, \ and\ \bibinfo {author} {\bibfnamefont {Y.~C.}\ \bibnamefont {San}},\ }\href@noop {} {\enquote {\bibinfo {title} {Anomaly detection in presence of irrelevant features},}\ } (\bibinfo {year} {2023}),\ \Eprint {http://arxiv.org/abs/2310.13057} {arXiv:2310.13057 [hep-ph]} \BibitemShut {NoStop}%
\bibitem [{\citenamefont {Buhmann}\ \emph {et~al.}(2023)\citenamefont {Buhmann}, \citenamefont {Ewen}, \citenamefont {Kasieczka}, \citenamefont {Mikuni}, \citenamefont {Nachman},\ and\ \citenamefont {Shih}}]{full_AD}%
  \BibitemOpen
  \bibfield  {author} {\bibinfo {author} {\bibfnamefont {E.}~\bibnamefont {Buhmann}}, \bibinfo {author} {\bibfnamefont {C.}~\bibnamefont {Ewen}}, \bibinfo {author} {\bibfnamefont {G.}~\bibnamefont {Kasieczka}}, \bibinfo {author} {\bibfnamefont {V.}~\bibnamefont {Mikuni}}, \bibinfo {author} {\bibfnamefont {B.}~\bibnamefont {Nachman}}, \ and\ \bibinfo {author} {\bibfnamefont {D.}~\bibnamefont {Shih}},\ }\href@noop {} {\enquote {\bibinfo {title} {Full phase space resonant anomaly detection},}\ } (\bibinfo {year} {2023}),\ \Eprint {http://arxiv.org/abs/2310.06897} {arXiv:2310.06897 [hep-ph]} \BibitemShut {NoStop}%
\bibitem [{\citenamefont {Kasieczka}\ \emph {et~al.}(2022)\citenamefont {Kasieczka}, \citenamefont {Nachman},\ and\ \citenamefont {Shih}}]{lhco}%
  \BibitemOpen
  \bibfield  {author} {\bibinfo {author} {\bibfnamefont {G.}~\bibnamefont {Kasieczka}}, \bibinfo {author} {\bibfnamefont {B.}~\bibnamefont {Nachman}}, \ and\ \bibinfo {author} {\bibfnamefont {D.}~\bibnamefont {Shih}},\ }\href {\doibase 10.5281/zenodo.6466204} {\enquote {\bibinfo {title} {{R\&D Dataset for LHC Olympics 2020 Anomaly Detection Challenge}},}\ } (\bibinfo {year} {2022})\BibitemShut {NoStop}%
\bibitem [{\citenamefont {Sjöstrand}\ \emph {et~al.}(2006)\citenamefont {Sjöstrand}, \citenamefont {Mrenna},\ and\ \citenamefont {Skands}}]{pythia_1}%
  \BibitemOpen
  \bibfield  {author} {\bibinfo {author} {\bibfnamefont {T.}~\bibnamefont {Sjöstrand}}, \bibinfo {author} {\bibfnamefont {S.}~\bibnamefont {Mrenna}}, \ and\ \bibinfo {author} {\bibfnamefont {P.}~\bibnamefont {Skands}},\ }\href {\doibase 10.1088/1126-6708/2006/05/026} {\bibfield  {journal} {\bibinfo  {journal} {Journal of High Energy Physics}\ }\textbf {\bibinfo {volume} {2006}},\ \bibinfo {pages} {026–026} (\bibinfo {year} {2006})}\BibitemShut {NoStop}%
\bibitem [{\citenamefont {Sjöstrand}\ \emph {et~al.}(2008)\citenamefont {Sjöstrand}, \citenamefont {Mrenna},\ and\ \citenamefont {Skands}}]{pythia_2}%
  \BibitemOpen
  \bibfield  {author} {\bibinfo {author} {\bibfnamefont {T.}~\bibnamefont {Sjöstrand}}, \bibinfo {author} {\bibfnamefont {S.}~\bibnamefont {Mrenna}}, \ and\ \bibinfo {author} {\bibfnamefont {P.}~\bibnamefont {Skands}},\ }\href {\doibase 10.1016/j.cpc.2008.01.036} {\bibfield  {journal} {\bibinfo  {journal} {Computer Physics Communications}\ }\textbf {\bibinfo {volume} {178}},\ \bibinfo {pages} {852–867} (\bibinfo {year} {2008})}\BibitemShut {NoStop}%
\bibitem [{\citenamefont {de~Favereau}\ \emph {et~al.}(2014)\citenamefont {de~Favereau}, \citenamefont {Delaere}, \citenamefont {Demin}, \citenamefont {Giammanco}, \citenamefont {Lemaître}, \citenamefont {Mertens},\ and\ \citenamefont {Selvaggi}}]{delphes_1}%
  \BibitemOpen
  \bibfield  {author} {\bibinfo {author} {\bibfnamefont {J.}~\bibnamefont {de~Favereau}}, \bibinfo {author} {\bibfnamefont {C.}~\bibnamefont {Delaere}}, \bibinfo {author} {\bibfnamefont {P.}~\bibnamefont {Demin}}, \bibinfo {author} {\bibfnamefont {A.}~\bibnamefont {Giammanco}}, \bibinfo {author} {\bibfnamefont {V.}~\bibnamefont {Lemaître}}, \bibinfo {author} {\bibfnamefont {A.}~\bibnamefont {Mertens}}, \ and\ \bibinfo {author} {\bibfnamefont {M.}~\bibnamefont {Selvaggi}},\ }\href {\doibase 10.1007/jhep02(2014)057} {\bibfield  {journal} {\bibinfo  {journal} {Journal of High Energy Physics}\ }\textbf {\bibinfo {volume} {2014}} (\bibinfo {year} {2014}),\ 10.1007/jhep02(2014)057}\BibitemShut {NoStop}%
\bibitem [{\citenamefont {Mertens}(2015)}]{delphes_2}%
  \BibitemOpen
  \bibfield  {author} {\bibinfo {author} {\bibfnamefont {A.}~\bibnamefont {Mertens}},\ }\href {\doibase 10.1088/1742-6596/608/1/012045} {\bibfield  {journal} {\bibinfo  {journal} {Journal of Physics: Conference Series}\ }\textbf {\bibinfo {volume} {608}},\ \bibinfo {pages} {012045} (\bibinfo {year} {2015})}\BibitemShut {NoStop}%
\bibitem [{\citenamefont {Selvaggi}(2014)}]{delphes_3}%
  \BibitemOpen
  \bibfield  {author} {\bibinfo {author} {\bibfnamefont {M.}~\bibnamefont {Selvaggi}},\ }\href {\doibase 10.1088/1742-6596/523/1/012033} {\bibfield  {journal} {\bibinfo  {journal} {Journal of Physics: Conference Series}\ }\textbf {\bibinfo {volume} {523}},\ \bibinfo {pages} {012033} (\bibinfo {year} {2014})}\BibitemShut {NoStop}%
\bibitem [{\citenamefont {Cacciari}\ and\ \citenamefont {Salam}(2006)}]{antikt_1}%
  \BibitemOpen
  \bibfield  {author} {\bibinfo {author} {\bibfnamefont {M.}~\bibnamefont {Cacciari}}\ and\ \bibinfo {author} {\bibfnamefont {G.~P.}\ \bibnamefont {Salam}},\ }\href {\doibase 10.1016/j.physletb.2006.08.037} {\bibfield  {journal} {\bibinfo  {journal} {Physics Letters B}\ }\textbf {\bibinfo {volume} {641}},\ \bibinfo {pages} {57–61} (\bibinfo {year} {2006})}\BibitemShut {NoStop}%
\bibitem [{\citenamefont {Cacciari}\ \emph {et~al.}(2008)\citenamefont {Cacciari}, \citenamefont {Salam},\ and\ \citenamefont {Soyez}}]{antikt_2}%
  \BibitemOpen
  \bibfield  {author} {\bibinfo {author} {\bibfnamefont {M.}~\bibnamefont {Cacciari}}, \bibinfo {author} {\bibfnamefont {G.~P.}\ \bibnamefont {Salam}}, \ and\ \bibinfo {author} {\bibfnamefont {G.}~\bibnamefont {Soyez}},\ }\href {\doibase 10.1088/1126-6708/2008/04/063} {\bibfield  {journal} {\bibinfo  {journal} {Journal of High Energy Physics}\ }\textbf {\bibinfo {volume} {2008}},\ \bibinfo {pages} {063–063} (\bibinfo {year} {2008})}\BibitemShut {NoStop}%
\bibitem [{\citenamefont {Cacciari}\ \emph {et~al.}(2012)\citenamefont {Cacciari}, \citenamefont {Salam},\ and\ \citenamefont {Soyez}}]{fastjet}%
  \BibitemOpen
  \bibfield  {author} {\bibinfo {author} {\bibfnamefont {M.}~\bibnamefont {Cacciari}}, \bibinfo {author} {\bibfnamefont {G.~P.}\ \bibnamefont {Salam}}, \ and\ \bibinfo {author} {\bibfnamefont {G.}~\bibnamefont {Soyez}},\ }\href {\doibase 10.1140/epjc/s10052-012-1896-2} {\bibfield  {journal} {\bibinfo  {journal} {The European Physical Journal C}\ }\textbf {\bibinfo {volume} {72}} (\bibinfo {year} {2012}),\ 10.1140/epjc/s10052-012-1896-2}\BibitemShut {NoStop}%
\bibitem [{\citenamefont {Thaler}\ and\ \citenamefont {Van~Tilburg}(2011)}]{nsub_1}%
  \BibitemOpen
  \bibfield  {author} {\bibinfo {author} {\bibfnamefont {J.}~\bibnamefont {Thaler}}\ and\ \bibinfo {author} {\bibfnamefont {K.}~\bibnamefont {Van~Tilburg}},\ }\href {\doibase 10.1007/jhep03(2011)015} {\bibfield  {journal} {\bibinfo  {journal} {Journal of High Energy Physics}\ }\textbf {\bibinfo {volume} {2011}} (\bibinfo {year} {2011}),\ 10.1007/jhep03(2011)015}\BibitemShut {NoStop}%
\bibitem [{\citenamefont {Thaler}\ and\ \citenamefont {Van~Tilburg}(2012)}]{nsub_2}%
  \BibitemOpen
  \bibfield  {author} {\bibinfo {author} {\bibfnamefont {J.}~\bibnamefont {Thaler}}\ and\ \bibinfo {author} {\bibfnamefont {K.}~\bibnamefont {Van~Tilburg}},\ }\href {\doibase 10.1007/jhep02(2012)093} {\bibfield  {journal} {\bibinfo  {journal} {Journal of High Energy Physics}\ }\textbf {\bibinfo {volume} {2012}} (\bibinfo {year} {2012}),\ 10.1007/jhep02(2012)093}\BibitemShut {NoStop}%
\bibitem [{\citenamefont {Shih}(2021)}]{extra_qcd}%
  \BibitemOpen
  \bibfield  {author} {\bibinfo {author} {\bibfnamefont {D.}~\bibnamefont {Shih}},\ }\href {\doibase 10.5281/zenodo.5759087} {\enquote {\bibinfo {title} {{Additional QCD Background Events for LHCO2020 R\&D (signal region only)}},}\ } (\bibinfo {year} {2021})\BibitemShut {NoStop}%
\bibitem [{\citenamefont {Papamakarios}\ \emph {et~al.}(2018)\citenamefont {Papamakarios}, \citenamefont {Pavlakou},\ and\ \citenamefont {Murray}}]{maf}%
  \BibitemOpen
  \bibfield  {author} {\bibinfo {author} {\bibfnamefont {G.}~\bibnamefont {Papamakarios}}, \bibinfo {author} {\bibfnamefont {T.}~\bibnamefont {Pavlakou}}, \ and\ \bibinfo {author} {\bibfnamefont {I.}~\bibnamefont {Murray}},\ }\href@noop {} {\enquote {\bibinfo {title} {Masked autoregressive flow for density estimation},}\ } (\bibinfo {year} {2018}),\ \Eprint {http://arxiv.org/abs/1705.07057} {arXiv:1705.07057 [stat.ML]} \BibitemShut {NoStop}%
\bibitem [{\citenamefont {Dinh}\ \emph {et~al.}(2015)\citenamefont {Dinh}, \citenamefont {Krueger},\ and\ \citenamefont {Bengio}}]{nice}%
  \BibitemOpen
  \bibfield  {author} {\bibinfo {author} {\bibfnamefont {L.}~\bibnamefont {Dinh}}, \bibinfo {author} {\bibfnamefont {D.}~\bibnamefont {Krueger}}, \ and\ \bibinfo {author} {\bibfnamefont {Y.}~\bibnamefont {Bengio}},\ }\href@noop {} {\enquote {\bibinfo {title} {Nice: Non-linear independent components estimation},}\ } (\bibinfo {year} {2015}),\ \Eprint {http://arxiv.org/abs/1410.8516} {arXiv:1410.8516 [cs.LG]} \BibitemShut {NoStop}%
\bibitem [{\citenamefont {Durkan}\ \emph {et~al.}(2019)\citenamefont {Durkan}, \citenamefont {Bekasov}, \citenamefont {Murray},\ and\ \citenamefont {Papamakarios}}]{rqs_paper}%
  \BibitemOpen
  \bibfield  {author} {\bibinfo {author} {\bibfnamefont {C.}~\bibnamefont {Durkan}}, \bibinfo {author} {\bibfnamefont {A.}~\bibnamefont {Bekasov}}, \bibinfo {author} {\bibfnamefont {I.}~\bibnamefont {Murray}}, \ and\ \bibinfo {author} {\bibfnamefont {G.}~\bibnamefont {Papamakarios}},\ }\href@noop {} {\enquote {\bibinfo {title} {Neural spline flows},}\ } (\bibinfo {year} {2019}),\ \Eprint {http://arxiv.org/abs/1906.04032} {arXiv:1906.04032 [stat.ML]} \BibitemShut {NoStop}%
\bibitem [{\citenamefont {Pedregosa}\ \emph {et~al.}(2011)\citenamefont {Pedregosa}, \citenamefont {Varoquaux}, \citenamefont {Gramfort}, \citenamefont {Michel}, \citenamefont {Thirion}, \citenamefont {Grisel}, \citenamefont {Blondel}, \citenamefont {Prettenhofer}, \citenamefont {Weiss}, \citenamefont {Dubourg}, \citenamefont {Vanderplas}, \citenamefont {Passos}, \citenamefont {Cournapeau}, \citenamefont {Brucher}, \citenamefont {Perrot},\ and\ \citenamefont {{{\'E}}douard Duchesnay}}]{scikit}%
  \BibitemOpen
  \bibfield  {author} {\bibinfo {author} {\bibfnamefont {F.}~\bibnamefont {Pedregosa}}, \bibinfo {author} {\bibfnamefont {G.}~\bibnamefont {Varoquaux}}, \bibinfo {author} {\bibfnamefont {A.}~\bibnamefont {Gramfort}}, \bibinfo {author} {\bibfnamefont {V.}~\bibnamefont {Michel}}, \bibinfo {author} {\bibfnamefont {B.}~\bibnamefont {Thirion}}, \bibinfo {author} {\bibfnamefont {O.}~\bibnamefont {Grisel}}, \bibinfo {author} {\bibfnamefont {M.}~\bibnamefont {Blondel}}, \bibinfo {author} {\bibfnamefont {P.}~\bibnamefont {Prettenhofer}}, \bibinfo {author} {\bibfnamefont {R.}~\bibnamefont {Weiss}}, \bibinfo {author} {\bibfnamefont {V.}~\bibnamefont {Dubourg}}, \bibinfo {author} {\bibfnamefont {J.}~\bibnamefont {Vanderplas}}, \bibinfo {author} {\bibfnamefont {A.}~\bibnamefont {Passos}}, \bibinfo {author} {\bibfnamefont {D.}~\bibnamefont {Cournapeau}}, \bibinfo {author} {\bibfnamefont {M.}~\bibnamefont {Brucher}}, \bibinfo {author} {\bibfnamefont {M.}~\bibnamefont {Perrot}}, \ and\ \bibinfo {author} {\bibnamefont
  {{{\'E}}douard Duchesnay}},\ }\href {http://jmlr.org/papers/v12/pedregosa11a.html} {\bibfield  {journal} {\bibinfo  {journal} {Journal of Machine Learning Research}\ }\textbf {\bibinfo {volume} {12}},\ \bibinfo {pages} {2825} (\bibinfo {year} {2011})}\BibitemShut {NoStop}%
\bibitem [{\citenamefont {Ke}\ \emph {et~al.}(2017)\citenamefont {Ke}, \citenamefont {Meng}, \citenamefont {Finley}, \citenamefont {Wang}, \citenamefont {Chen}, \citenamefont {Ma}, \citenamefont {Ye},\ and\ \citenamefont {Liu}}]{lightgbm}%
  \BibitemOpen
  \bibfield  {author} {\bibinfo {author} {\bibfnamefont {G.}~\bibnamefont {Ke}}, \bibinfo {author} {\bibfnamefont {Q.}~\bibnamefont {Meng}}, \bibinfo {author} {\bibfnamefont {T.}~\bibnamefont {Finley}}, \bibinfo {author} {\bibfnamefont {T.}~\bibnamefont {Wang}}, \bibinfo {author} {\bibfnamefont {W.}~\bibnamefont {Chen}}, \bibinfo {author} {\bibfnamefont {W.}~\bibnamefont {Ma}}, \bibinfo {author} {\bibfnamefont {Q.}~\bibnamefont {Ye}}, \ and\ \bibinfo {author} {\bibfnamefont {T.-Y.}\ \bibnamefont {Liu}},\ }in\ \href {https://proceedings.neurips.cc/paper_files/paper/2017/file/6449f44a102fde848669bdd9eb6b76fa-Paper.pdf} {\emph {\bibinfo {booktitle} {Advances in Neural Information Processing Systems}}},\ Vol.~\bibinfo {volume} {30},\ \bibinfo {editor} {edited by\ \bibinfo {editor} {\bibfnamefont {I.}~\bibnamefont {Guyon}}, \bibinfo {editor} {\bibfnamefont {U.~V.}\ \bibnamefont {Luxburg}}, \bibinfo {editor} {\bibfnamefont {S.}~\bibnamefont {Bengio}}, \bibinfo {editor} {\bibfnamefont {H.}~\bibnamefont {Wallach}},
  \bibinfo {editor} {\bibfnamefont {R.}~\bibnamefont {Fergus}}, \bibinfo {editor} {\bibfnamefont {S.}~\bibnamefont {Vishwanathan}}, \ and\ \bibinfo {editor} {\bibfnamefont {R.}~\bibnamefont {Garnett}}}\ (\bibinfo  {publisher} {Curran Associates, Inc.},\ \bibinfo {year} {2017})\BibitemShut {NoStop}%
\bibitem [{\citenamefont {Paszke}\ \emph {et~al.}(2019)\citenamefont {Paszke}, \citenamefont {Gross}, \citenamefont {Massa}, \citenamefont {Lerer}, \citenamefont {Bradbury}, \citenamefont {Chanan}, \citenamefont {Killeen}, \citenamefont {Lin}, \citenamefont {Gimelshein}, \citenamefont {Antiga}, \citenamefont {Desmaison}, \citenamefont {Köpf}, \citenamefont {Yang}, \citenamefont {DeVito}, \citenamefont {Raison}, \citenamefont {Tejani}, \citenamefont {Chilamkurthy}, \citenamefont {Steiner}, \citenamefont {Fang}, \citenamefont {Bai},\ and\ \citenamefont {Chintala}}]{pytorch}%
  \BibitemOpen
  \bibfield  {author} {\bibinfo {author} {\bibfnamefont {A.}~\bibnamefont {Paszke}}, \bibinfo {author} {\bibfnamefont {S.}~\bibnamefont {Gross}}, \bibinfo {author} {\bibfnamefont {F.}~\bibnamefont {Massa}}, \bibinfo {author} {\bibfnamefont {A.}~\bibnamefont {Lerer}}, \bibinfo {author} {\bibfnamefont {J.}~\bibnamefont {Bradbury}}, \bibinfo {author} {\bibfnamefont {G.}~\bibnamefont {Chanan}}, \bibinfo {author} {\bibfnamefont {T.}~\bibnamefont {Killeen}}, \bibinfo {author} {\bibfnamefont {Z.}~\bibnamefont {Lin}}, \bibinfo {author} {\bibfnamefont {N.}~\bibnamefont {Gimelshein}}, \bibinfo {author} {\bibfnamefont {L.}~\bibnamefont {Antiga}}, \bibinfo {author} {\bibfnamefont {A.}~\bibnamefont {Desmaison}}, \bibinfo {author} {\bibfnamefont {A.}~\bibnamefont {Köpf}}, \bibinfo {author} {\bibfnamefont {E.}~\bibnamefont {Yang}}, \bibinfo {author} {\bibfnamefont {Z.}~\bibnamefont {DeVito}}, \bibinfo {author} {\bibfnamefont {M.}~\bibnamefont {Raison}}, \bibinfo {author} {\bibfnamefont {A.}~\bibnamefont {Tejani}}, \bibinfo
  {author} {\bibfnamefont {S.}~\bibnamefont {Chilamkurthy}}, \bibinfo {author} {\bibfnamefont {B.}~\bibnamefont {Steiner}}, \bibinfo {author} {\bibfnamefont {L.}~\bibnamefont {Fang}}, \bibinfo {author} {\bibfnamefont {J.}~\bibnamefont {Bai}}, \ and\ \bibinfo {author} {\bibfnamefont {S.}~\bibnamefont {Chintala}},\ }\href@noop {} {\enquote {\bibinfo {title} {Pytorch: An imperative style, high-performance deep learning library},}\ } (\bibinfo {year} {2019}),\ \Eprint {http://arxiv.org/abs/1912.01703} {arXiv:1912.01703 [cs.LG]} \BibitemShut {NoStop}%
\bibitem [{\citenamefont {Kingma}\ and\ \citenamefont {Ba}(2017)}]{Adam}%
  \BibitemOpen
  \bibfield  {author} {\bibinfo {author} {\bibfnamefont {D.~P.}\ \bibnamefont {Kingma}}\ and\ \bibinfo {author} {\bibfnamefont {J.}~\bibnamefont {Ba}},\ }\href@noop {} {\enquote {\bibinfo {title} {Adam: A method for stochastic optimization},}\ } (\bibinfo {year} {2017}),\ \Eprint {http://arxiv.org/abs/1412.6980} {arXiv:1412.6980 [cs.LG]} \BibitemShut {NoStop}%
\bibitem [{\citenamefont {Loshchilov}\ and\ \citenamefont {Hutter}(2019)}]{adamw}%
  \BibitemOpen
  \bibfield  {author} {\bibinfo {author} {\bibfnamefont {I.}~\bibnamefont {Loshchilov}}\ and\ \bibinfo {author} {\bibfnamefont {F.}~\bibnamefont {Hutter}},\ }\href@noop {} {\enquote {\bibinfo {title} {Decoupled weight decay regularization},}\ } (\bibinfo {year} {2019}),\ \Eprint {http://arxiv.org/abs/1711.05101} {arXiv:1711.05101 [cs.LG]} \BibitemShut {NoStop}%
\end{thebibliography}%

\end{document}